\documentclass[a4paper,aps,10pt]{revtex4}
\usepackage{epsf,amssymb,amsmath,latexsym,theorem}
\usepackage{graphicx,verbatim,color}

\renewcommand{\theequation}{\arabic{section}.\arabic{equation}}
\def \slas{\kern -6.2pt /}
\def \sla{\kern -5.4pt /}
\def \sl{\kern -4.0pt /}
\def \Cslas{\kern -6.8pt /}
\def \Dslas{\kern -6.8pt /}
\def \slass{\kern -7.4pt /}

\def \ii{{\mathrm{i}}}
\def \d{{\mathrm{d}}}

\def \lcdi{\tilde{\di}}
\def \lcx{\tilde{x}}

\def \qb{\bar q}      % q bar
\def \Tr{\text{Tr}}  % Tr
\def \di{\mathbf{d}} % d (interior d)
\def \D{{\cal D}}      

\begin{document}
\title{$B$-Meson Distribution Amplitudes of Geometric Twist vs.~Dynamical Twist}
\author{Bodo Geyer} 
\email{geyer@itp.uni-leipzig.de}
\author{Oliver Witzel\footnote{Present address: Humboldt Universit\"at zu Berlin, 
Institut f\"ur Physik, Newtonstr. 15, 12489 Berlin, Germany}}
\email{witzel@physik.hu-berlin.de}
\affiliation{Universit\"at Leipzig, Institut f\"ur Theoretische Physik, 
Augustusplatz 10, 04109 Leipzig, Germany}

\date{\today}

\begin{abstract}
Two- and three-particle distribution amplitudes of heavy pseudoscalar mesons of well-defined geo\-metric twist are introduced. They are obtained from appropriately parametrized  vacuum-to-meson matrix elements by applying those twist projectors which determine the enclosed light-cone operators of definite geometric twist and, in addition, observing the heavy quark constraint.  %
Comparing these distribution amplitudes with the conventional ones of dynamical twist we derive relations between them, partially being of Wandzura-Wilczek type; also sum rules of Burkhardt-Cottingham type are derived.
The derivation is performed for the (double) Mellin moments and then re-summed to the non-local distribution amplitudes.
Furthermore, a parametrization of vacuum-to-meson matrix elements 
for non-local operators off the light-cone in terms of distribution amplitudes accompanying independent kinematical structures is derived. 
\end{abstract}

\maketitle

\vspace*{-0.5cm}
\section{Introduction}
\setcounter{equation}{0}

Exclusive non-leptonic decays of $B$ mesons play a crucial role for our understanding 
of rare flavour-changing processes and the exploration of the mechanism of $CP$ violation 
within the standard model. Thereby, of special interest are hadronic two-body decays,
either via $B \rightarrow D$ transition, e.g.,~$B \rightarrow D\pi\,(D K)$, or with two 
very energetic light mesons in the final state, e.g.,~$B \rightarrow \pi\pi\,(K\pi)$. 
While the weak interaction part of these processes is fairly well understood 
their strong interaction dynamics is quite non-trivial. However, some 
simplifications are possible due to the strong ordering of the three fundamental 
scales, the weak interaction scale $M_W$, the $b$-quark mass $m_b$, and the QCD 
scale $\Lambda_{\rm QCD}$. Because $m_b \gg \Lambda_{\rm QCD}$ 
the heavy quark effective theory (HQET) \cite{HQET,FGGW} (for a review, see, Refs.~\cite{HQETRev}) may be applied and, furthermore, the strong interaction effects with 
virtualities above $m_b$ may be included into the renormalized coefficients of 
local operators ${\cal O}_i$ of the weak effective Hamiltonian. 

In order to compute the (renormalized) matrix elements 
$\langle M_1 M_2 | {\cal O}_i | B \rangle$, at least in leading order of
$\Lambda_{\rm QCD}/m_b$, some factorization \cite{WSB} into perturbatively calculable 
short-distance contributions and appropriate long-distance contributions 
has to be applied -- either using 
the $QCD$-factorization approach \cite{Beneke2000,BF01,QCDfact}, 
the more effective $SCET$ approach \cite{SCET,BPRS2004}, or the 
$pQCD$ approach \cite{KLS}. For example, according to \cite{Beneke2000} that
matrix element in  case of two light mesons can be represented by hard scattering 
amplitudes $T$,  
$(BM)$ form factors $F^{BM}_j$ as well as light ($\Phi_{M}$) and heavy 
($\Phi_{B}$) meson light-cone (LC) distribution amplitudes (DA), e.g.,
\begin{align}
\hspace{-.2cm}
\langle M_1 M_2 | {\cal O}_i | \bar B \rangle
=&
\sum_j F^{BM_1}_j(m_2^2) \int_0^1\! du\,T^I_{ij}(u)\,\Phi_{M_2}(u) 
%~+~ (M_1 \longleftrightarrow M_2) \nonumber\\ &&
+
\int_0^1\! d\xi\,du\,dv\,T^{II}_i(\xi,u,v)\,\Phi_{B}(\xi)\,\Phi_{M_1}(v)\,\Phi_{M_2}(u),
\end{align}
assuming $M_1$ to pick up the spectator quark from the $B$-meson; obviously, no 
long-distance interaction takes place between $M_2$ and the $(BM_1)$-system.
 
In the case of light (pseudo)scalar and vector mesons the light-cone distribution 
amplitudes (LCDA) are well-known for leading and next-to-leading twist for 
bilocal (quark-antiquark) as well as trilocal (quark-gluon-antiquark)operators for the $\pi$-meson \cite{Chernyak1984,Braun1990} and the $\rho$-meseon
\cite{Ball1996,Ball1998}, 
also considering the Wandzura-Wilczek (WW) relation. 
In  case of B-mesons they have been determined in the framework of HQET \cite{Grozin1996,KPY00,BF01} also discussing in detail the WW-approximation 
\cite{Kodaira2001,HWZ05}; the case of the $D$-meson is easily obtained, at least in 
leading order, by observing the spin-flavour symmetry of HQET.
Furthermore, with the aim of a better understanding of the scale dependence of LCDA 
-- and of the hard scattering kernel -- in the factorization procedure the
knowledge of their renormalization behaviour is required. In the case of the 
leading LCDA $\Phi_+$ this has been studied recently \cite{REN}.

In the limit of infinite heavy quark masses $(m_Q\to\infty)$, the heavy quark field 
$Q(x)$ reduces to an \emph{effective} field $h_v(x)$ with the kinematics contained 
in a phase factor ($v$ being the heavy meson's velocity, $v^2=1$),
\begin{equation}
Q(x)\to e^{-\ii m_Q v x}h_v(x). 
\label{Qfield}
\end{equation}
Moreover, the effective heavy quark field $h_v(x)$ obeys the {\em on-shell constraint},
\begin{equation}
v\sla h_v = h_v ,
\label{on-shell} 
\end{equation} 
and the effective Lagrangian 
$ %\begin{equation}
{\cal L}_\text{eff}= \sum_{i=1}^{N_h}\bar h_v^{(i)} \ii (vD) h_v^{(i)},
$ %\end{equation}
is independent of
the spin or mass of the heavy quark and exhibits, therefore, the $U(2N_h)$ spin-flavor 
symmetry ($N_h$ is the number of contributing heavy flavors). Of course, the on-shell 
constraint reduces the number of independent heavy meson DAs in comparision with 
light mesons.

In accordance with the definition of usual meson LCDAs \cite{Chernyak1984} 
but additionally respecting the on-shell constraint (\ref{on-shell}), the 
$B$-meson LCDAs arise by parametrizing matrix elements of appropriate non-local 
LC operators ${\cal O}_i$ which are built up by quark and antiquark fields 
-- occasionally containing also gluons and/or quark-antiquark pairs -- 
sandwiched between vacuum $\langle 0|$ and $B$-meson state $|B(v)\rangle$ of 
momentum $P = M\,v$. For instance, the (2-particle) LCDAs are introduced 
as (phase factor omitted)
\begin{equation} 
\label{me}
\langle 0| \bar q(\kappa_1 \lcx)\, \Gamma\, U(\kappa_1 \lcx,\kappa_2 \lcx)\,
h_v(\kappa_2 \lcx) | B(v) \rangle = {\mathcal K}_\Gamma(P,x) 
\int_{-1}^1 \d\xi\; \varphi_B(\xi)\; e^{-\ii(\kappa_1-\kappa_2)\,(\lcx P)\, \xi},
\end{equation}
where $\Gamma$ denotes some generic Dirac structure 
\begin{equation}
\label{gamma}
\Gamma = \{1,\gamma_\alpha, \ii \sigma_{\alpha\beta},
\gamma_5, \gamma_5\gamma_\alpha, \ii \gamma_5 \sigma_{\alpha\beta}\}
\quad \text{with} \quad \sigma_{\alpha\beta} =
\hbox{\large$ \frac{\ii}{2}$}[\gamma_\alpha,\gamma_\beta],
\end{equation}
and
$ %\begin{align}
\lcx = x + v\big( \sqrt{(vx)^2-x^2} - (vx) \big),~  \lcx^2 =0, 
$ %\end{align}
defines some light-ray being related to $x$ by a fixed non-null subsidiary four-vector 
which may be identified with the $B$-meson's velocity. The path ordered phase factor,
$ U(\kappa_1 \lcx, \kappa_2 \lcx)={\cal P} \exp\big\{-ig
\int^{\kappa_1}_{\kappa_2}d\tau \; \lcx^\mu A_\mu(\tau \lcx)\big\},$ 
assuming Schwinger-Fock gauge, will be omitted in the following. 
Furthermore, the matrix element (\ref{me})
is parametrized by a kinematic factor ${\mathcal K}_\Gamma$ and the Fourier transform
of the DA $\varphi_B(\xi)$ w.r.t.~variable $\lcx P$; ${\mathcal K}_\Gamma$ depends
on the momentum $P$ of the (pseudoscalar) meson and LC coordinate $\lcx$ of the nonlocal  operator as well as on the generic Dirac structure $\Gamma$. Explicit forms for 
${\mathcal K}_\Gamma$ will be introduced in Section \ref{parametrization}.
Everywhere, possible color indices in operator matrix elements will be suppressed.
The integration range in (\ref{me}) results from the fact that, in the framework of
non-local LC expansion \cite{Anikin1978}, that matrix element can be shown to 
be an entire analytic function in the variable $\lcx P$ \cite{Geyer1994}. Usually, 
due to the (anti)symmetry of the relevant QCD operators 
${\cal O}_\Gamma(\kappa_1 \lcx, \kappa_2 \lcx)$ w.r.t. exchange 
$\kappa_1 \leftrightarrow \kappa_2$ the integration range is restricted to 
$0 \le \xi \le 1$. ---
Analogous definitions hold in the case of tri-local operators including, e.g., the
gluon field strength $F_{\mu\nu}(\kappa\lcx)$ at arbitrary intermediate points 
$\kappa\in [\kappa_1,\kappa_2]$.

Conventionally, LCDAs are characterized by its {\em dynamical twist} which, roughly speaking, 
counts powers of $M/Q$ for the various terms in the kinematic decomposition 
of the matrix elements of non-local QCD operators \cite{Jaffe1992}. Alternatively, 
using group theoretical arguments, the original definition of twist 
\cite{Gross1971} for local QCD operators, $\tau = $ dimension $d - $ (Lorentz) spin $j$, 
has been generalized to the notion of {\em geometric twist} for non-local 
QCD tensor operators on the light-cone \cite{Geyer1999,Geyer2000b,Lazar2002} as well 
as off the light-cone \cite{Eilers2003,Joerg}. The decomposition of such tensor 
operators into operators of definite geometric twist leads to corresponding 
decompositions of the LCDAs \cite{Geyer2001,Geyer2000} and to their power 
(or target mass) corrections \cite{GRE04}.

In fact, concerning phenomenological aspects the notion of dynamical twist is favored. 
But, from a quantum field theoretical point of view, geometrical twist seems to
be more appropriate since it has an obvious group theoretical meaning and, therefore,
should have well-defined renormalization properties; it also offers a clear
separation between radiative corrections and higher (geometric) twist effects. 

Both definitions of twist, despite of being different for higher twist,
coincide at leading twist. However, by comparing equivalent kinematical structures, 
it has been shown for distribution functions in DIS and of LCDAs for light mesons, 
especially for $\pi$- and $\rho$-mesons, that there exist {\em unique relationships between 
the distributions of geometrical twist and the usual ones of dynamical twist}. 

To be more specific, let us generically denote the distributions of definite 
geometric twist $\tau$ by $\varphi_i^{(\tau)}(\eta)$ and the ones of 
dynamical twist $t$ by $\phi_j^{(t)}(\xi)$. Then the distributions of given
dynamical twist $t$ are uniquely determined by that set of distributions of geometric
twist $\tau$ with $\tau_{\rm min} \leq \tau \leq t$ and,
vice versa:
\begin{align}
\phi_j^{(t)}(\xi) = 
\sum_{\tau = \tau_{\rm min}}^t \int \d \eta\, 
{K_{j}}^{i}(\xi,\eta)\,\varphi_i^{(\tau)}(\eta),
\qquad %\nonumber\\
\varphi_i^{(\tau)}(\eta) = 
\sum_{t = t_{\rm min}}^\tau \int \d \xi\, 
{(K^{-1})_i}^j(\eta,\xi)\,\phi_j^{(t)}(\xi)\,,
\label{triangle}
\end{align}
with some invertible kernel ${K_{j}}^{i}(\xi,\eta)$, where 
$t_{\rm min}= \tau_{\rm min}$. 
In fact, the relations between both kinds of distributions are 
of triangular shape and, therefore, the corresponding set of equations 
can be solved with respect to either basis. {\em Solving the distributions 
of dynamical twist w.r.t.~those of geometric twist allows to 
derive the well-known WW-relations} together with additional WW-like
relations \cite{Geyer2000}; corresponding relations have been derived 
for the (light) meson LCDAs \cite{Lazar2000} and, later on, called 
{\em geometrical WW-relations} \cite{BL01}.  Using 
in addition the equations of motion, a different set of WW-like relations 
for the LCDAs -- especially for the vector meson case -- appeared 
\cite{Ball1998}; they should be called {\em dynamical WW-relations}. 
This situation has been discussed more detailed in Refs.~\cite{BL01}.
 
In the case of heavy meson LCDAs the situation suffers from the on-shell
constraint of HQET. First, the number of independent LCDAs is reduced. They
have been determined already by Grozin and Neubert \cite{Grozin1996} and 
discussed further by various authors \cite{KPY00,BF01,Kodaira2001,HWZ05}, 
also considering the WW-relation within the dynamical twist approach. 
Second, concerning renormalization, they show some pecularities which have
been studied more detailed in the case of leading $B$-meson LCDA in Refs.~\cite{REN}. 
Also regarding this it seems to be of interest to {\em consider the geometric 
twist approach, too, and to find how these two approaches are related}. 
The results of such a comparison, mainly based on Ref.~\cite{W04}, will be 
presented here. 

The paper is organized as follows. 
In Sect.~\ref{parametrization} we briefly repeat the derivation of the non-local
two- and three-particle LCDAs of dynamical twist by using the trace formalism
according to Refs.~\cite{BF01,Kodaira2001,L05}; in addition, in order to
be able to compare with the corresponding LCDAs of geometric twist, their local
form are given as Mellin moments. A consistent parametrization of the relevant  
matrix elements which is not restricted to the light-cone is given in 
Appendix~\ref{GenAnsatz}.
In Sect.~\ref{ParaGeom} we determine the local two- and three-particle LCDAs of 
geometrical twist by applying the local projection operators (restricted to its
light-cone form) onto the matrix elements of the corresponding non-local LC
operators. In order not to be confused by the on-shell constraint and obstacles
of renormalization, they are derived for the general case also applying to light 
(pseudo)scalar mesons. The corresponding projection operators are well known 
from earlier work up to tensor operators of second rank \cite{Geyer2000,Joerg}; 
their local form is given in  Appendix~\ref{projectors}.
In Sect.~\ref{relations} the relations between the LCDAs of definite dynamical 
and geometrical twist are given, first, for their local form and then for their
non-local form. Thereby, we also derive the relations between LCDAs of definite
geometric twist resulting from the on-shell constraint. Most of the calculations 
are performed using FORM \cite{FORM}.

%%%\hspace*{2.0cm}
%%%\newpage
%%%\input{HeavyQuarks}
%%%\hspace*{2.0cm}
%%%\newpage
%%%\input{twist}
%%%\hspace*{2.0cm}
%%%\newpage
%\newpage
\setcounter{equation}{0} 
\section{Distribution amplitudes of dynamical twist}
\label{parametrization}

To begin with we briefly review the conventional representation of vacuum-to-meson 
matrix elements of bi- and trilocal {\em light-cone} operators for $B$-mesons in the 
heavy quark limit by LCDAs of dynamical twist times corresponding kinematic structures. 
Hereby we follow the covariant trace-formalism \cite{FGGW}.
These LCDAs are Fourier transformed and converted into their Mellin moments.

%The relevance of introducing DAs of dynamical twist is given 
%by the fact that, phenomenologically, these amplitudes are related to powers of $1/Q$ 
%which are counted as {\em dynamical} twist. 
%
%In case of bilocal operators, we choose dimensionless DAs multiplied 
%by the meson mass $M$, whereas, in case of trilocal operators, the DAs 
%carry dimension two and are multiplied by $M$ times $\lcx$. 

In the trace-formalism the vacuum-to-meson transition by a generic quark-antiquark
operator with a single heavy quark is parametrized in terms of two-particle LCDAs 
$\hat\Phi_\pm(v\lcx)$ as follows \cite{BF01,Kodaira2001}:
\begin{align}
\label{Kodaira_bi}
\langle 0| \qb(\lcx) \Gamma h_v(0)|B(v) \rangle
~=~-\frac{\ii f_B M}{2}\;\text{Tr}\left\{ \gamma_5\,\Gamma \frac{1+v\sla}{2} 
\left(\hat \Phi_+(v\lcx) - \frac{\lcx\sla}{2(v\lcx)}
\left[\hat\Phi_+(v\lcx)-\hat\Phi_-(v\lcx)\right]\right)\right\},
\end{align}
where, as usual, the $B$-meson decay constant is defined by
$ %\begin{align}
\langle 0 |\qb(0)\gamma_\alpha\gamma_5  h_v(0) | B(v) \rangle~=~ \ii f_B M v_\alpha, %\label{decay_p}
$ %\end{align}
and $M$ is the mass of the $B$-meson. 
Computing these traces for the various Dirac structures $\Gamma$ one obtains:
\begin{align}
\langle 0|\qb(\lcx)&  \gamma_5 h_v(0)|B(v) \rangle
~=~-\ii f_B M \;\hbox{\Large$\frac{1}{2}$} \left[\hat \Phi_+ +\hat \Phi_-\right] ,
\label{ff1}\\
\langle 0|\qb(\lcx)& \gamma_5 \gamma_\alpha h_v(0)|B(v) \rangle
~=~-\ii f_B M \left( v_\alpha \Phi_+ - \frac{\lcx_\alpha}{2(v\lcx)}%\frac{1}{2} 
\left[\hat \Phi_+ - \hat \Phi_-\right]  \right),
\label{ff2}\\
\langle 0|\qb(\lcx)& \gamma_5 \ii \sigma_{\alpha\beta} h_v(0)|B(v) \rangle~=~\ii f_B M \;
\frac{v_{[\alpha} \lcx_{\beta]}}{(v\lcx)}%\frac{1}{2}
\left[\hat\Phi_+ -\hat  \Phi_- \right],
\label{ff3} 
\end{align}
with arguments $v\lcx$ omitted. The parametrization of the matrix element 
$\langle 0|\qb(\lcx) \ii \sigma_{\alpha\beta}h_v(0)|B(v) \rangle$ 
is obtained simply by observing the relation $\ii \sigma_{\alpha\beta}= 
(-\ii/2) \epsilon_{\alpha\beta\kappa\lambda}\gamma_5 \ii \sigma^{\kappa\lambda}$, 
whereas the matrix elements 
$\langle 0|\qb(\lcx) \ii \sigma_{\alpha\beta}h_v(0)|B(v) \rangle$ and
$\langle 0|\qb(\lcx) h_v(0)|B(v) \rangle $ vanish.
Here, and in the following, we use the notation
\begin{align}
a_{[\alpha}b_{\beta]} := \left(a_{\alpha}b_{\beta} - a_{\beta}b_{\alpha}\right)/2,
\qquad
a_{(\alpha}b_{\beta)} := \left(a_{\alpha}b_{\beta} + a_{\beta}b_{\alpha}\right)/2.
\label{bracket}
\end{align}
 
As Grozin and Neubert \cite{Grozin1996} state, in the limit of fast-moving mesons,
$\hat\Phi_+$ is of leading (dynamical) twist while $\hat\Phi_-$ is sub-leading.
When the matrix element (\ref{ff2}) is considered it becomes obvious that $v_\alpha$
is related to leading twist while $\lcx_\alpha/(v\lcx)$ should correspond to $1/Q$ and 
therefore is related to subleading twist.

Equivalently, the vacuum-to-meson matrix element containing a trilocal quark-antiquark-gluon operator is parametrized in terms of four three-particle LCDAs $\hat\Psi_A(v\lcx;\vartheta),\,\hat\Psi_V(v\lcx;\vartheta),\,\hat X_A(v\lcx;\vartheta)$ 
and $\hat Y_A(v\lcx;\vartheta)$ with $\vartheta$ being restricted to $0\leq\vartheta\leq 1$ 
as follows \cite{Kodaira2001,L05}:
\begin{align}
\langle 0 | \qb(\lcx) F_{\mu\nu}(\vartheta\lcx) \lcx^\nu \Gamma h_v(0) | B(v)\rangle~&=~\frac{\ii f_B M}{2} \;
\Tr \bigg\{\gamma_5 \Gamma \frac{1+v\sla}{2}
\bigg( \left(v_\mu \lcx\sla - (v\lcx)\gamma_\mu\right) \left[\hat\Psi_A- \hat\Psi_V\right](v\lcx;\vartheta) 
\nonumber\\
&\hspace{1.9cm}
 -\ii \sigma_{\mu\nu} \lcx^\nu \hat\Psi_V(v\lcx;\vartheta) 
 - \lcx_\mu \hat X_A(v\lcx;\vartheta) 
 + \frac{\lcx_\mu\lcx\sla}{(v\lcx)}\hat Y_A(v\lcx;\vartheta) \bigg) \bigg\}. 
\label{Kodaira_tri}
\end{align}
Again, computing these traces for the Dirac structures $\Gamma$ 
(arguments $v\lcx$ and $\vartheta$ omitted) one obtains:
\begin{align}
 \langle 0|\qb(\lcx)&F_{\mu\nu}(\vartheta \lcx)\lcx^\nu\gamma_5 h_v(0)|B(v) 
\rangle~=~- \ii f_B M\,
\lcx_\mu\left[\hat X_A-\hat Y_A\right],
\label{TRI1}\\
 \langle 0|\qb(\lcx)&F_{\mu\nu}(\vartheta \lcx) \lcx^\nu \gamma_5 \gamma_\alpha h_v(0)|B(v) \rangle~=~\ii f_B M 
\bigg\{ \left(v_\mu \lcx_\alpha -(v\lcx) g_{\mu\alpha} \right) \hat \Psi_A
- v_\alpha \lcx_\mu\, \hat X_A + \frac{\lcx_\mu \lcx_\alpha}{(v\lcx)} \;
\hat Y_A \bigg\}, 
\label{TRI2}\\
%\intertext{}%
\langle 0|\qb(\lcx)&F_{\mu\nu}(\vartheta \lcx)\lcx^\nu \gamma_5 \ii \sigma_{\alpha\beta} h_v(0)|B(v) \rangle 
\nonumber \\ 
&\qquad ~=~2\,\ii f_B M 
\bigg\{ g_{\mu[\alpha} \lcx_{\beta]} \,\hat\Psi_V
- \frac{\lcx_\mu}{(v\lcx)} \lcx_{[\alpha} v_{\beta]} \,\hat Y_A
+\Big(\!v_\mu v_{[\alpha} \lcx_{\beta]} 
+(v\lcx) g_{\mu[\alpha} v_{\beta]} \! \Big)\!
\left[\hat \Psi_A -\hat\Psi_V\right]\!\!\bigg\},
\label{TRI3}\\
\langle 0|\qb(\lcx)&F_{\mu\nu}(\vartheta \lcx)\lcx^\nu \gamma_\alpha
h_v(0)|B(v) \rangle~=~- f_B M \;
\epsilon_{\mu\alpha\kappa\lambda} \,v^\kappa \lcx^\lambda \,\hat \Psi_V.
\phantom{\bigg\{\bigg\}}
\label{TRI6}
\end{align}

Again, the parametrization of the matrix element 
$\langle 0|\qb(\lcx)F_{\mu\nu}(\vartheta \lcx)\lcx^\nu \ii \sigma_{\alpha\beta}
h_v(0)|B(v) \rangle$ follows trivially from Eq.~(\ref{TRI3}), and 
$\langle 0|\qb(\lcx)F_{\mu\nu}(\vartheta \lcx)\lcx^\nu h_v(0)|B(v) \rangle$
vanishes. 
According to the above introduced conventions, $\Psi_A$, $\Psi_V$ and $X_A$ are of leading (dynamical) twist, whereas $Y_A$ is of subleading twist. 

An independent derivation of the just introduced parametizations, but not restricted 
to the light-cone, is presented in  Appendix \ref{GenAnsatz}. In this more general 
case the matrix elements depend on three leading DAs, $X_1-X_5, X_2$ and $X_4$, and three subleading ones, $Y_2, Y_4$ and $Y_5$, (cf. Eq.~(\ref{R0})). There it is also shown, that
{\em off the light-cone}  $\Psi_A$ and $\Psi_V$ contain also subleading contributions,
and $\Phi_-$ contains subleading contributions already {\em on the light-cone}!

Later on, to obtain relations between  DAs of dynamical and geometric twist, the DAs 
have to be Fourier-transformed and, thereafter, be converted into Mellin moments. 
As has been argued in Refs.~\cite{Grozin1996,Kodaira2001} the singularities of the DAs in the 
complex $(v\lcx)$ plane are such that their Fourier transforms $\Phi_\pm(u)$ vanish 
for $u<0$. On the other
hand, considering non-forward matrix elements of light-cone operators, it has been 
shown with the help of the $\alpha$-parameter representation of Feynman diagrams, 
that these matrix elements are entire analytic functions with respect to $(P\lcx)$ and,
thus, the support of $\Phi_\pm(u)$ is restricted to $-1\le u \le +1$ \cite{Geyer1994,Geyer2001}.  

For LCDAs which are related to bilocal operators their Fourier transforms and
the corresponding Mellin moments read
\begin{align}
\hat \Phi_\pm(v\lcx)~&=~\int_0^1\!\! \d u\;e^{-\ii u P\lcx} \Phi_\pm(u)
=~ \sum_{n=0}^\infty \frac{(-\ii P\lcx)^n}{n!} \Phi_{\pm|n} ,\\
%\intertext{where are defined by}
\Phi_{\pm|n}~&=~\int_0^1\!\!\d u\; u^n \;\Phi_{\pm}(u). 
\label{MellinMoments}
\end{align}
In case of trilocal operators two-parameter LCDAs occur whose Fourier transforms 
and the corresponding double Mellin moments read
\begin{align}
\hat F(v\lcx, \vartheta v\lcx)~
&=~\int_0^1 \!\!\D u_i\; e^{-\ii(u_1+\vartheta u_2)P\lcx}\; F(u_i)
=~ \sum_{n=0}^\infty \frac{(-\ii P\lcx)^n}{n!} F_{n}(\vartheta),
%\sum_{m=0}^n \binom{n}{m}\; \vartheta^m\; F_{n,m},
\\
F_{n}(\vartheta) &= \sum_{m=0}^n \binom{n}{m}\; \vartheta^m\; F_{n,m},
\qquad
F_{n,m}~=~\int_0^1\!\!\D u_i\; u_1^{n-m} u_2^m \;F(u_i); 
\label{DoubleMoments}
\end{align}
here $\int_0^1 \!\!\D u_i\;=\int_0^1 \!\!\d u_1\int_0^1 \!\!\d u_2$ 
and $\hat F(v\lcx;\vartheta)$ generically denotes any three-particle DA.

After transformation, Eqs.~(\ref{ff1}) -- (\ref{ff3}) read 
\begin{align}
\label{K1}
\langle 0 |\qb(\lcx)& \gamma_5 h_v(0) | B(v) \rangle~
%=~-  \ii f_B M  \int_0^1\!\! \d u \;\hbox{\Large$\frac{1}{2}$} 
%\left[ \Phi_+ + \Phi_-\right](u)\; e^{-\ii u P\lcx}, \\&\hspace{2.4cm}
=~- \ii f_B M \sum_{n=0}^\infty\frac{(-\ii  P\lcx)^n}{n!}
 \;\hbox{\Large$\frac{1}{2}$} \left[ \Phi_{+|n} + \Phi_{-|n} \right] ,
% \label{K1n}
 \\
\langle 0 |\qb(\lcx)&\gamma_5 \gamma_\alpha h_v(0) | B(v) \rangle
% ~=~- \ii f_B M \int_0^1\!\! \d u \; \Big( v_\alpha \Phi_+(u)
%- \frac{\lcx_\alpha}{2(v\lcx)}\left[\Phi_+ - \Phi_-\right](u)\Big)\;e^{-\ii u P\lcx},
%\\&\hspace{2.4cm}
 ~=~- \ii f_B M\sum_{n=0}^\infty\frac{(-\ii  P\lcx)^n}{n!} \Big( v_\alpha \Phi_{+|n}
 - \frac{\lcx_\alpha}{2(v\lcx)}\left[\Phi_{+|n}- \Phi_{-|n}\right]\Big) , 
 \label{K2n}\\
\langle 0 |\qb(\lcx)&\gamma_5 \ii \sigma_{\alpha\beta} h_v(0) | B(v) \rangle
% ~=~\ii f_B M \,\int_0^1\!\! \d u \;\frac{ v_\alpha \lcx_\beta - \lcx_\alpha v_\beta }{2(v\lcx)}
% \left[\Phi_+ - \Phi_-\right](u)\;e^{-\ii u P\lcx},\\&\hspace{2.4cm}
 ~=~\ii f_B M\sum_{n=0}^\infty\frac{(-\ii  P\lcx)^n}{n!} 
 \frac{ v_{[\alpha} \lcx_{\beta]}}{(v\lcx)}
 \left[\Phi_{+|n}- \Phi_{-|n}\right],
 \label{K3n}
\end{align}
while, the transformed equations, in case of a matrix element of a 
trilocal operator, read
\begin{align}
\langle 0|\qb(\lcx)& F_{\mu\nu}(\vartheta \lcx)\lcx^\nu \gamma_5 h_v(0)|B(v) \rangle
%~=~-\ii f_B M \int_0^1\!\!\D u_i \; \lcx_\mu \left[X_A - Y_A\right](u_i)
%\;e^{-\ii (u_1+\vartheta u_2) P\lcx},\\
=~- \ii f_B M\, \lcx_\mu \sum_{n=0}^\infty \frac{(-\ii P\lcx)^{n}}{n!} 
%\sum_{m=0}^n \binom{n}{m} \vartheta^m % \frac{}{(v\lcx)}
\left[X_{A|{n}}(\vartheta) - Y_{A|{n}}(\vartheta)\right],
\label{K4n}\\
%\end{align}
\langle 0|\qb(\lcx)& F_{\mu\nu}(\vartheta \lcx) \lcx^\nu \gamma_5 \gamma_\alpha h_v(0)|B(v) \rangle~
\nonumber \\
%=&~\ii f_B M \int_0^1\!\!\D u_i\;\bigg\{ \Big(v_\mu \lcx_\alpha
%- g_{\mu\alpha}(v\lcx) \Big) \Psi_A(u_i) 
%- \lcx_\mu v_\alpha X_A(u_i) + \frac{\lcx_\mu \lcx_\alpha}{(v\lcx)}Y_A(u_i) \bigg\}
%\; e^{-\ii (u_1+\vartheta u_2) P\lcx}, \\
=&~\ii f_B M \sum_{n=0}^\infty\frac{(-\ii P\lcx)^{n}}{n!}
%\sum_{m=0}^n \binom{n}{m} \vartheta^m 
 \left\{ \left(v_\mu \lcx_\alpha
- (v\lcx)\, g_{\mu\alpha} \right)\Psi_{A|{n}}(\vartheta)
- \lcx_\mu v_\alpha X_{A|{n}}(\vartheta)
+ \frac{\lcx_\mu \lcx_\alpha}{(v\lcx)}Y_{A|{n}} (\vartheta)\right\}, 
\label{K5n}\\
%\end{align}
%\begin{align}
\langle 0|\qb(\lcx)&F_{\mu\nu}(\vartheta \lcx) \lcx^\nu \gamma_5 \ii \sigma_{\alpha\beta} h_v(0)|B(v) \rangle~
\nonumber \\
%=&~\ii f_B M \int_0^1\!\!\D u_i\; \bigg\{  \Big(v_\mu v_{[\alpha} \lcx_{\beta]}
% +  g_{\mu[\alpha}v_{\beta]}(v\lcx) \Big) \left[ \Psi_A - \Psi_V \right](u_i)  
% \nonumber \\&\qquad\qquad\qquad\qquad\qquad\qquad\qquad
%+  g_{\mu[\alpha} \lcx_{\beta]}\;\Psi_V(u_i)
%- \frac{\lcx_{\mu}}{(v\lcx)} \lcx_{[\alpha} v_{\beta]} \; Y_A(u_i)  \bigg\} 
%\;e^{-\ii (u_1+\vartheta u_2) P\lcx},\\
=&~2\,\ii f_B M\sum_{n=0}^\infty \frac{(-\ii P\lcx)^{n}}{n!}
%\sum_{m=0}^n \binom{n}{m} \vartheta^m
 \bigg\{  \left(v_\mu v_{[\alpha} \lcx_{\beta]} 
 + (v\lcx)\, g_{\mu[\alpha}v_{\beta]}  \right)
 \left[ \Psi_{A|{n}}(\vartheta) - \Psi_{V|{n}}(\vartheta) \right]  
\nonumber \\&\qquad\qquad\qquad\qquad\qquad\qquad\qquad
+   g_{\mu[\alpha} \lcx_{\beta]}\; \Psi_{V|{n}}(\vartheta)
-\lcx_\mu \frac{\lcx_{[\alpha} v_{\beta]} }{(v\lcx)} \;
Y_{A|{n}}(\vartheta)  \bigg\}  ,
\label{K6n}\\
%
%\end{align}
%\begin{align}
\langle 0|\qb(\lcx)& F_{\mu\nu}(\vartheta \lcx)\lcx^\nu \gamma_\alpha h_v(0)|B(v) \rangle
%~=~- f_B M \;
%\epsilon_{\mu\alpha\kappa\lambda} 
%\int_0^1\!\!\D u_i\; v^\kappa \lcx^\lambda \Psi_V(u_i) \;e^{-\ii (u_1+\vartheta u_2) P\lcx}, \\
~=~- f_B M \, \epsilon_{\mu\alpha\kappa\lambda}\, v^\kappa \lcx^\lambda
\sum_{n=0}^\infty\frac{(-\ii  P\lcx)^{n}}{n!} 
%\sum_{m=0}^n \binom{n}{m} \vartheta^m
 \Psi_{V|{n}}(\vartheta).
\label{K2}
\end{align}
The local decompositions (\ref{K1}) -- (\ref{K2}) w.r.t.~dynamical twist 
are required for comparison with those of geometric twist in Sect.~\ref{relations}.

%\newpage
\setcounter{equation}{0} 
\section{Distribution amplitudes of geometric twist}
\label{ParaGeom}

In this Section we introduce the $B$-meson LCDAs of definite {\em geometric twist}. 
This is done in analogy to the introduction of quark distribution functions in deep 
inelastic scattering \cite{Geyer2000b} and of $\rho$-meson LCDAs \cite{Lazar2000}
in terms of definite geometric twist. Namely, we use the decomposition of non-local light-cone operators ${\cal O}_{\{\sigma\}}$ with given tensor structure $\{\sigma\}$
into a (finite) sum of non-local tensor operators of definite twist $\tau$, 
\begin{align}
{\cal O}_{\{\sigma\}} &= \sum_{\tau} {\cal O}^{(\tau)}_{\{\sigma\}}
\qquad {\rm with} \qquad
{\cal O}^{(\tau)}_{\{\sigma\}} = 
\widetilde{\cal P}_{\{\sigma\}}^{(\tau)\,\{\sigma'\}} {\cal O}_{\{\sigma'\}},
\intertext{with appropriate projection operators,
$\widetilde{\cal P}_{\{\sigma\}}^{(\tau)\,\{\sigma'\}}\equiv
\widetilde{\cal P}_{\{\sigma\}}^{(\tau)\,\{\sigma'\}}(\lcx,\lcdi)$,
already known from Refs.~\cite{Geyer1999,Geyer2000,Joerg},}
\widetilde{\cal P}_{\{\sigma\}}^{(\tau)\,\{\sigma'\}} &= ~
\left(\widetilde{\cal P}^{(\tau)}\,, 
~
\widetilde{\cal P}_{\alpha}^{(\tau)\alpha'}\,, 
~
\widetilde{\cal P}_{[\alpha\beta]}^{(\tau)\,[\alpha'\beta']} \,,
~
\widetilde{\cal P}_{(\alpha\beta)}^{(\tau)(\alpha'\beta')}\,,
~\ldots ~ \right)\,;
\label{pro}
\end{align}
$\lcdi$ is the inner derivative on the light-cone (see, Appendix \ref{projectors}).
Obviously, for a given tensor structure, the sum over these projection operators of 
different twist $\tau$ defines a decomposition of unity, 
$\sum_{\tau} \widetilde{\cal P}_{\{\sigma\}}^{(\tau)\,\{\sigma'\}}
= \delta_{\{\sigma\}}^{\{\sigma'\}}$. 

Considering bilocal operators, the corresponding meson LCDAs of definite geometric 
twist $\tau$, generically denoted by $\varphi^{(\tau)}_a(u)$, are introduced 
according to Ref.~\cite{Geyer2001} (cf.~also~Refs.~\cite{Geyer2000b,Lazar2000})
\begin{align}
\langle 0|{\cal O}_{\{\sigma\}}(\lcx, 0)|B(v) \rangle
= \sum_{\tau} \widetilde{\cal P}_{\{\sigma\}}^{(\tau)\,\{\sigma'\}}(\lcx,\lcdi)\,
{\cal K}^{[s]a}_{\{\sigma'\}}(v,\lcx)
\int_0^1\!\! \d u\;e^{-\ii u P\lcx} \varphi^{(\tau)}_a(u).
\label{BNL}
\end{align}
Thereby ${\cal K}^{[s]a}_{\{\sigma'\}}(v,\lcx)$ is the {\em basic kinematical 
structure} (of scale dimension $s$ w.r.t.~$x\partial$) of that matrix element which 
can be read off from its parametrization w.r.t.~LCDAs of \emph{dynamical} twist since 
at leading order geometric and dynamical twist coincide by construction.
When, in accordance with (\ref{MellinMoments}), one goes over to Mellin moments 
$\varphi^{(\tau)}_{a|n}$ one has to apply the corresponding local projection operators 
$\widetilde{\cal P}_{\{\sigma\}|n+s}^{(\tau)\,\{\sigma'\}}(\lcx,\lcdi)$, 
\begin{align}
\langle 0|{\cal O}_{\{\sigma'\}}(\lcx, 0)|B(v) \rangle
&= \sum_{\tau} \sum_{n=0}^\infty
 \widetilde{\cal P}_{\{\sigma\}|n+s}^{(\tau)\,\{\sigma'\}}(\lcx,\lcdi)\,
{\cal K}^{[s]a}_{\{\sigma'\}}(v,\lcx)
\frac{(-\ii P\lcx)^n}{n!} \varphi^{(\tau)}_{a|n};
\label{BL}
\end{align}
in case of trilocal matrix elements 
$\langle 0|{\cal O}_{\{\sigma'\}}(\lcx,\vartheta\lcx, 0)|B(v) \rangle$
the moment $\varphi^{(\tau)}_{a|n}$ has to be replaced by
$\Upsilon^{(\tau)}_{a | n}(\vartheta)$.

The explicit form of the local projection operators is given in the
Appendix \ref{projectors}. In the following they will be applied to the bi- 
and tri-local vacuum-to-meson matrix elements for pseudoscalar mesons. 
At first, without applying the heavy quark on-shell constraint. Thereby 
we obtain a decomposition of these matrix elements which differs from the 
decomposition in terms of dynamical twist determined in the preceding Section. 
Afterwards, we will resum the Mellin moments to get the corresponding $B$-meson DAs.

\subsection{Twist decomposition: Local representation in terms of Mellin moments}

First, we consider the bi-local operators and introduce the Mellin moments
of the corresponding DAs of definite geometric twist. Then, applying the local twist 
projectors, in particular (\ref{vec_0}) -- (\ref{vec_2}) in the (axial)vector case
 and (\ref{ten_A1}) -- (\ref{ten_A3}) in the skew tensor case,  
 we get the following decomposition of the vacuum-to-meson matrix elements:
\begin{align}
\langle 0|\qb(\lcx) \gamma_5 h_v(0)|B(v) \rangle
&~=~\ii f_B M \;\sum_\tau
 \sum_{n=0}^\infty \widetilde {\cal P}^{(\tau)}_n
 \frac{(-\ii  P\lcx)^n}{n!}\;\varphi_{P|n}^{(\tau)} ,
 \nonumber\\
 \label{Dbi_s}
&~=~\ii f_B M 
 \sum_{n=0}^\infty \frac{(-\ii  P\lcx)^n}{n!}\;\varphi_{P|n}^{(3)} ,
 \\
\langle 0|\qb(\lcx)\gamma_5 \gamma_\alpha h_v(0)|B(v) \rangle
&~=~\ii f_B M\;\sum_\tau
\sum_{n=0}^\infty  \widetilde{\cal P}^{(\tau)\alpha^\prime}_{\alpha|n}
\frac{(-\ii  P\lcx)^n}{n!}\; v_{\alpha^\prime} \varphi_{A|n}^{(\tau)},
\nonumber\\
 \label{Dbi_v}
&~=~\ii f_B M
\sum_{n=0}^\infty\frac{(-\ii  P\lcx)^n}{n!} \left\{ v_\alpha \varphi_{A|n}^{(2)} 
- \frac{\lcx_\alpha}{2(v\lcx)}\frac{n}{n+1}
\left[ \varphi_{A|n}^{(2)} -\varphi_{A|n}^{(4)}\right] \right\} ,
\\
\langle 0|\qb(\lcx) \gamma_5 \ii \sigma_{\alpha\beta} h_v(0)|B(v) \rangle
&~=~\ii f_B M \;\sum_\tau
\sum_{n=0}^\infty \widetilde{\cal P}^{(\tau)[\alpha^\prime\beta^\prime]}_{[\alpha\beta]|n} 
\frac{(-\ii  P\lcx)^n}{n!}
\frac{v_{[\alpha^\prime} \lcx_{\beta^\prime]}}{(v\lcx)} 
\; \varphi_{T|n}^{(\tau)} 
\nonumber\\
\label{Dbi_a}
&~=~\ii f_B M 
\sum_{n=1}^\infty \frac{(-\ii  P\lcx)^n}{n!} 
\frac{v_{[\alpha} \lcx_{\beta]}}{(v\lcx)}
\;\varphi_{T|n}^{(3)} .
\end{align}
The (pseudo)scalar case is trivial since on the light-cone always
 $\widetilde {\cal P}^{(\tau_{\rm min})} = 1$. Therefore only the single LCDA 
 $\varphi_{P|n}^{(3)}$ occurs.
Obviously, in the axial vector and skew tensor case some LCDAs do not appear, namely,
$\varphi_{A|n}^{(3)}$ in Eq.~(\ref{Dbi_v}) and $\varphi_{T|n}^{(2)}$ as well as 
$\varphi_{T|n}^{(4)}$ in Eq.~(\ref{Dbi_a}) vanish. The summation in Eq.~(\ref{Dbi_a})
begins at $n=1$ since $\varphi_{T|0}^{(3)}$ vanishes identically (cf. Eq.~(\ref{ten_A2})).
The matrix elements of scalar and 
vector operators vanish completely, and that with $\ii\sigma_{\alpha\beta}$ results 
trivially from expression (\ref{Dbi_a}). 
As stated above, the twist decomposition generates all linearly independent kinematic coefficients which are known from the previous Section, Eqs.~(\ref{ff1}) -- (\ref{ff3}),
as well as Eqs.~(\ref{S9}) -- (\ref{S11}) of  Appendix \ref{GenAnsatz}.
Let us also remark that, apart from different
normalization, the leading LCDAs from (\ref{Dbi_s}) -- (\ref{Dbi_a}) coincide with 
those of Ref.~\cite{BF01}, whereby the additional contribution in (\ref{Dbi_v})
is  considered to be nonleading (w.r.t.~dynamical twist).
\medskip

Next, we consider the trilocal operators and introduce the double Mellin moments
of corresponding three-particle LCDAs of definite geometric twist. Concerning the
twist projections, we remind that only the tensorial structure of the operators 
is crucial and not if it is a bilocal or trilocal one. 
The relevant projection operators are defined by (\ref{vec_0}) -- (\ref{vec_2})
for the (axial)vector operators, by (\ref{ten_A1}) -- (\ref{ten_A3}) for the skew tensor 
operators and by (\ref{ten_S0}) -- (\ref{ten_S4}) for the symmetric tensor operators of 
second rank. Unfortunately, at present no projectors for tensors of third rank, besides totally symmetric ones, are at our disposal. 

With these operators the trilocal vacuum-to-meson matrix elements in case of a 
pseudoscalar meson read:
\begin{align}
\langle 0|\qb(\lcx)&F_{\mu\nu}(\vartheta \lcx)\lcx^\nu \gamma_5 h_v(0)|B(v)\rangle
\nonumber\\
\label{tri_v}
=&~\ii f_B M\;\sum_\tau
\sum_{n=0}^\infty %\sum_{m=0}^n \binom{n}{m} \vartheta^m 
 \widetilde{\cal P}^{(\tau)\mu^\prime}_{\mu|n+1}
 \frac{(-\ii  P\lcx)^{n}}{n!} \lcx_{\mu^\prime} \Upsilon_{P|{n}}^{(\tau)}(\vartheta)
 \\
 \label{Dtri_v}
=&~ \ii f_B M \lcx_\mu
\sum_{n=0}^\infty \frac{(-\ii  P\lcx)^{n}}{n!} 
%\sum_{m=0}^n \binom{n}{m} \vartheta^m 
\Upsilon_{P|{n}}^{(5)}(\vartheta)\,,\\
%
%\intertext{}
%
\langle 0|\qb(\lcx)&F_{\mu\nu}(\vartheta \lcx) \lcx^\nu \gamma_5 \gamma_\alpha
h_v(0)|B(v) \rangle~\nonumber\\
=&~-\ii f_B M \;\sum_\tau 
\sum_{n=0}^\infty  %\sum_{m=0}^n \binom{n}{m} \vartheta^m 
\bigg\{
\widetilde{\cal P}_{[\mu\alpha]|n+1}^{(\tau)[\mu^\prime\alpha^\prime]}
\frac{(-\ii  P\lcx)^{n}}{n!} 
\lcx_{[\mu^\prime} v_{\alpha^\prime]} 
\left[\Upsilon_{A1|{n}}^{(\tau)}(\vartheta)+\Upsilon_{A2|{n}}^{(\tau)}(\vartheta)\right]
\nonumber \\
&\qquad
+~ %\sum_{n=0}^\infty  
\widetilde{\cal P}_{(\mu\alpha)|n+1}^{(\tau)(\mu^\prime\alpha^\prime)}
\frac{(-\ii  P\lcx)^{n}}{n!}  
%\sum_{m=0}^n \binom{n}{m} \vartheta^m
\bigg( g_{\mu^\prime\alpha^\prime} (v\lcx)\Upsilon_{A1|{n}}^{(\tau)}(\vartheta)
- \, \lcx_{(\mu^\prime} v_{\alpha^\prime)}\! %+v_{\mu^\prime} \lcx_{\alpha^\prime}\right) 
\left[ \Upsilon_{A1|{n}}^{(\tau)}(\vartheta) - \Upsilon_{A2|{n}}^{(\tau)}(\vartheta)\right]\!\bigg)\!\bigg\}
\label{tri_t2}\\
\label{Dtri_t2}
=&~ \ii f_B M 
\sum_{n=0}^\infty \frac{(-\ii P\lcx)^{n}}{n!}
\bigg\{  %\!%\sum_{m=0}^n \binom{n}{m} \vartheta^m
\bigg(\! \Big(v_\mu \lcx_\alpha-(v\lcx)\, g_{\mu\alpha} \Big) \Upsilon_{A1|{n}}^{(4)}(\vartheta)
- \lcx_\mu v_\alpha\! \Upsilon_{A2|{n}}^{(4)}(\vartheta)\!\bigg)
\nonumber \\ 
& \qquad\qquad\qquad\qquad
- \frac{\lcx_\mu \lcx_\alpha}{2(v\lcx)}
%\sum_{n=1}^\infty 
%\frac{(-\ii  P\lcx)^{n}}{n!}%\!%\sum_{m=0}^n \binom{n}{m} \vartheta^m
 \frac{n}{n+1}\bigg(\!
  \left[\Upsilon_{A1|{n}}^{(4)}(\vartheta) - \Upsilon_{A1|{n}}^{(6)}(\vartheta)\right] 
- \left[\Upsilon_{A2|{n}}^{(4)}(\vartheta) - \Upsilon_{A2|{n}}^{(6)}(\vartheta)\right] \!\bigg)\!\bigg\},
\\
%
%\intertext{}
%
\langle 0|\qb(\lcx)&F_{\mu\nu}(\vartheta \lcx)\lcx^\nu \gamma_\alpha h_v(0)|B(v) \rangle
\nonumber\\
=&~- f_B M \sum_\tau
 \sum_{n=0}^\infty %\sum_{m=0}^n \!\!\binom{n}{m} \vartheta^m
 \widetilde{\cal P}^{(\tau)[\mu^\prime\alpha^\prime]}_{[\mu\alpha]|n+1}
 \frac{(-\ii  P\lcx)^{n}}{n!}\, \epsilon_{\mu^\prime\alpha^\prime\kappa\lambda}
 v^\kappa \lcx^\lambda \Upsilon_{V|{n}}^{(\tau)}(\vartheta)
\label{tri_t2d}\\
\label{Dtri_t2d}
=&~- f_B M\;
\epsilon_{\mu\alpha\kappa\lambda}\,v^\kappa \lcx^\lambda
\sum_{n=1}^\infty \frac{(-\ii  P\lcx)^{n}}{n!} %\sum_{m=0}^n \binom{n}{m} \vartheta^m
\Upsilon_{V|{n}}^{(4)}(\vartheta)\,,\\
%\intertext{} 
% 
\langle 0|\qb(\lcx)&F_{\mu\nu}(\vartheta \lcx)\lcx^\nu
\gamma_5 \ii \sigma_{\alpha\beta} h_v(0)|B(v) \rangle
\nonumber\\
=&~2\,\ii f_B M\;\sum_\tau
\sum_{n=0}^\infty %\sum_{m=0}^n \binom{n}{m} \vartheta^m 
\widetilde{\cal P}^{(\tau)(\mu^\prime[\alpha^\prime)\beta^\prime]}_{(\mu[\alpha)\beta]|n+1} 
\frac{(-\ii  P\lcx)^{n}}{n!} 
 \Big(v_{\mu^\prime}\,v_{[\alpha^\prime} \lcx_{\beta^\prime]}
     + g_{\mu^\prime[\alpha^\prime} v_{\beta^\prime]} (v\lcx) \Big) 
\Upsilon_{T|{n}}^{(\tau)}(\vartheta) \,.
\label{tri_t3}
\end{align}

Here, some remarks are in order: 

First, since the field strength $F_{\mu\nu}$ in the trilocal operators (\ref{tri_v}) 
-- (\ref{tri_t3}) is contracted with $\lcx^\nu$ and, consequently, the kinematic 
terms ${\cal K}^a(v,\lcx)$, according to relations (\ref{K4n}) -- (\ref{K2}), 
have scale dimension 1, the local LC projection operators are to be taken for $n+1$. 
Let us remind that in Schwinger-Fock gauge, $\lcx^\nu A_\nu(\lcx) =0$, the 
field strength is related to the gauge potential, 
$\kappa\lcx^\nu F_{\mu\nu}(\kappa \lcx) 
= -\, (1+\kappa \partial/\partial\kappa) A_\mu(\kappa \lcx)$.

Second, although any tensor of second rank can be split into a symmetric and an 
antisymmetric part, it is impossible to yield an input parametrization for the 
matrix element in (\ref{tri_t2}) with only one set of LCDAs 
$\Upsilon_{As|{n}}^{(\tau)}(\vartheta)$ associated with the symmetric 
and another set $\Upsilon_{Aa|{n}}^{(\tau)}(\vartheta)$ associated with the 
antisymmetric coefficients. The reason is, that the input parametrization has 
to vanish if additionally contracted with $\lcx^\mu$. Therefore, both LCDAs
interfere for the $g_{\mu\alpha}$-term. 

Third, as mentioned above, concerning expression (\ref{tri_t3}) we do not know the 
explicit structure of the projection operator  
$\widetilde{\cal P}^{(\tau)(\mu^\prime[\alpha^\prime)\beta^\prime]}_{(\mu[\alpha)\beta]|n}$.
Furthermore, a projection operator 
$\widetilde{\cal P}^{(\tau)[\mu^\prime\alpha^\prime\beta^\prime]}_{[\mu\alpha\beta]|n}$
does not occur since an $\epsilon$-structure on the RHS is forbidden for pseudoscalar 
mesons.

Finally, looking at the expressions (\ref{Dtri_v}), (\ref{Dtri_t2}) and (\ref{Dtri_t2d}) 
we observe again
that a huge number of LCDAs of definite twist vanishes, thereby having in mind, that an additional free index $\mu$ besides $\Gamma$ comes into play. In the pseudoscalar 
case only the highest twist part $\Upsilon_{P|{n}}^{(5)}$ and in the vector case only
$\Upsilon_{V|{n}}^{(4)}$ occur (also here $\Upsilon_{V|{0}}^{(4)}\equiv 0 $); 
in the axial vector case, despite of being more complicated, 
only the LCDAs $\Upsilon_{Ai|{n}}^{(4)}$ and $\Upsilon_{Ai|{n}}^{(6)}$ occur.
Since the on-shell constraint reduces the number of independent DAs by two, we suppose
that in the skew tensor case only two additional independent DAs may occur. 

\medskip

Independently, there occurs another possible set of trilocal vacuum-to-meson matrix 
elements and their corresponding LCDAs, denoted by $\Omega$, which are related to the three-particle operators
$\qb(\lcx) F_{\mu\nu}(\vartheta \lcx) \gamma^\mu \lcx^\nu \Gamma h_v(0)$. 
According to their tensor structure they are analogously defined as the bilocal ones,
Eqs.~(\ref{Dbi_s}) -- (\ref{Dbi_a}). Therefore, we note only their form in terms of Mellin moments as follows:
\begin{align}
\label{Dtri_g_s}
 \langle 0|\qb(\lcx)& F_{\mu\nu}(\vartheta \lcx)\gamma^\mu \lcx^\nu  \gamma_5
h_v(0)|B(v) \rangle~=~\ii f_B M (v\lcx)
\sum_{n=0}^\infty  \frac{(-\ii  P\lcx)^n}{n!} %\sum_{m=0}^n \binom{n}{m} \vartheta^m
\Omega_{P|{n}}^{(4)}(\vartheta)\,,\\
\label{Dtri_g_v}
 \langle 0|\qb(\lcx)& F_{\mu\nu}(\vartheta \lcx) \gamma^\mu \lcx^\nu
\gamma_5 \gamma_\alpha h_v(0)|B(v) \rangle
 ~=~\ii f_B M (v\lcx)
\sum_{n=0}^\infty \frac{(-\ii  P\lcx)^n}{n!} %\sum_{m=0}^n \binom{n}{m} \vartheta^m 
\nonumber \\ &\hspace{5cm}
\times \Bigg\{  v_\alpha \Omega_{A|{n}}^{(3)}(\vartheta) 
-\frac{\lcx_\alpha}{2(v\lcx)} 
\frac{n}{n+1}\left[ \Omega_{A|{n}}^{(3)}(\vartheta) 
- \Omega_{A|{n}}^{(5)}(\vartheta) \right]
\Bigg\} ,\\
\label{Dtri_g_a}
\langle 0|\qb(\lcx)& F_{\mu\nu}(\vartheta \lcx)\gamma^\mu \lcx^\nu 
\gamma_5 \ii \sigma_{\alpha\beta} h_v(0)|B(v) \rangle~=~2\, \ii\, f_B M (v\lcx)
\sum_{n=1}^\infty \frac{(-\ii  P\lcx)^n}{n!} %\sum_{m=0}^n \binom{n}{m} \vartheta^m
 \frac{v_{[\alpha} \lcx_{\beta]}}{(v\lcx)}
\Omega_{T|{n}}^{(4)}(\vartheta)\, .
\end{align}

Obviously, the matrix elements of equations (\ref{Dtri_g_s}) -- (\ref{Dtri_g_a}) are 
related to those of equations (\ref{tri_v}) -- (\ref{tri_t3}) by identities of 
Dirac matrices. In particular, these relations read
\begin{align}
 \langle 0|\qb(\lcx)& F_{\mu\nu}(\vartheta \lcx) \gamma^\mu \lcx^\nu
\gamma_5 \gamma_\alpha h_v(0)|B(v) \rangle
~=~
- \langle 0|\qb(\lcx) F_{\alpha\nu}(\vartheta \lcx) \lcx^\nu
\gamma_5 h_v(0)|B(v) \rangle 
\nonumber \\ 
&\qquad\qquad\qquad\qquad\qquad\qquad\qquad\quad~
+g^{\lambda\mu}\langle 0|\qb(\lcx) F_{\mu\nu}(\vartheta \lcx) \lcx^\nu\gamma_5\ii 
\sigma_{\lambda\alpha} h_v(0)|B(v) \rangle,
\label{luck}\\
 \langle 0|\qb(\lcx)& F_{\mu\nu}(\vartheta \lcx)\gamma^\mu \lcx^\nu  \gamma_5
h_v(0)|B(P) \rangle
~=~-\,g^{\lambda\mu}
\langle 0|\qb(\lcx) F_{\mu\nu}(\vartheta \lcx)\lcx^\nu  
\gamma_5 \gamma_\lambda h_v(0)|B(v) \rangle,
\label{fix}\\
\langle 0|\qb(\lcx)& F_{\mu\nu}(\vartheta \lcx)\gamma^\mu \lcx^\nu
\gamma_5 \ii \sigma_{\alpha\beta} h_v(0)|B(v) \rangle
~=~
\left(g^{\mu}_{\phantom{\mu}\alpha} g^{\lambda}_{\phantom{\lambda}\beta} 
     -g^{\mu}_{\phantom{\mu}\beta} g^{\lambda}_{\phantom{\lambda}\alpha}\right)
\langle 0|\qb(\lcx) F_{\mu\nu}(\vartheta \lcx)\lcx^\nu  
\gamma_5 \gamma_\lambda h_v(0)|B(v) \rangle 
\nonumber \\&\qquad\qquad\qquad\qquad\qquad\qquad\qquad\qquad
+\ii \epsilon^{\mu\lambda}_{\phantom{\mu\lambda}\alpha\beta}\,
\langle 0|\qb(\lcx) F_{\mu\nu}(\vartheta \lcx)\lcx^\nu
\gamma_\lambda h_v(0)|B(v) \rangle.
\label{fax}
\end{align}
This leads to relations for the corresponding LCDAs which allows to express the 
$\Omega$'s via the $\Upsilon$'s. From Eqs.~(\ref{fix}) and (\ref{fax}) we receive 
the following two relations,
%valid for $n\ge 1$:
\begin{align}
\Omega_{P|{n}}^{(4)}(\vartheta)~
&\equiv~3\,\Upsilon_{A1|{n}}^{(4)}(\vartheta) +\Upsilon_{A2|{n}}^{(4)}(\vartheta)\, ,
\label{oh1}\\
\Omega_{T|n}^{(4)}(\vartheta)~
&\equiv~
2\,\Upsilon_{V|{n}}^{(4)}(\vartheta) +
\Upsilon_{A1|{n}}^{(4)}(\vartheta) + \Upsilon_{A2|{n}}^{(4)}(\vartheta), 
\label{oh2}
\end{align}
corresponding to expressions $\Upsilon_1$ and $\Upsilon_2$ in Eq.~(\ref{Upsilon}).
However, since we cannot manage the tensor of third rank in Eq.~(\ref{luck}), 
two further relations concerning $\Omega_{A|{n}}^{(3)}(\vartheta)$ and
$\Omega_{A|{n}}^{(5)}(\vartheta)$ are missing.

The matrix elements containing $F_{\mu\nu}(\vartheta \lcx)v^\mu \lcx^\nu$ 
follow immediately from the expressions (\ref{tri_v}) -- (\ref{tri_t3}) 
by multiplication with $v^\mu$. Regarding this we should remark that,
contrary to the expressions just derived, the corresponding result has {\em not} 
the same structure as it would follow from the bilocal operator!

%Yet, all the DAs which have been derived in this Subsection up to now do not obey the 
%constraints required to hold in the heavy quark limit. The DAs hence should obey 
%additional relations reducing the number of linearly independent DAs. This will be
%considered in detail in the next Section.
%
%On the one hand, 
%the DAs of each kind of meson, e.g.~a matrix element of a bilocal operator for a 
%pseudoscalar meson, obey mutual relations due to the on-shell constraint. 

\subsection{Twist decomposition: Nonlocal representation in terms of distribution amplitudes}

From the local results, we yield the corresponding nonlocal representation by going back to integral expressions. The fractions in $n$ transform thereby to a second integral according to
\begin{align}
 \frac{1}{n-r+1}\; \psi_n~
 &=~\int_0^1\!\d u\; u^n \int_u^1 \frac{\d w}{w}  
 \left(\frac{w}{u} \right)^r \psi(w),
 \label{Mellin1} 
% \\
% -\frac{1}{(n-r+1)^2}\; \psi_n~
% &=~\int_0^1\!\d u\; u^n \int_u^1 \frac{\d w}{w}  
% \left(\frac{w}{u} \right)^r \ln \left(\frac{u}{w} \right) \psi(w).
% \label{Mellin2}
\end{align}
Moreover, we have to respect that not all summations include the zeroth moment. 
Rewriting such sums by the exponentials minus the missing moment we
use the following formula:
\begin{align}
\int_0^1\!\d u\; \Big( e^{-\ii u P\lcx}-1 \Big)\psi(u) 
 &= \int_0^1\!\d u\,  e^{-\ii u P\lcx}\left( \psi(u)
 -\int_u^1 \frac{\d w}{w}\; \delta\left(\frac{u}{w} \right) \psi(w)\right).
% ,\\
% \int_0^1\!\d u\; \Big( e^{-\ii u P\lcx}-1-\left(-\ii u P\lcx\right) \Big)\psi(u)  
% &= \int_0^1\!\d u\, e^{-\ii u P\lcx} 
% \int_u^1 \frac{\d w}{w}\,\left(\frac{u}{w}+1 \right)(-\ii w P\lcx)^2\psi(w).
\end{align}
We thereby yield expressions all multiplied by the same exponential which will be essential, later on, for comparison with the distribution amplitudes of dynamical twist.

The twist-decomposed two-particle distribution amplitudes read in the nonlocal representation 
\begin{align}
\label{Ibi_s}
\langle 0|\qb(\lcx)& \gamma_5 h_v(0)|B(v) \rangle
 ~=~ \ii f_B M\int_0^1\!\d u\; \varphi_{P}^{(3)}(u) \; e^{-\ii u P\lcx},
\\
\label{Ibi_v}
\langle 0|\qb(\lcx)& \gamma_5 \gamma_\alpha h_v(0)|B(v) \rangle
 ~=~ \ii f_B M\int_0^1\!\d u \; \bigg\{ v_\alpha \;\varphi_{A}^{(2)}(u) 
 \nonumber\\
&\qquad\qquad\qquad\qquad\quad 
- \frac{\lcx_\alpha}{2(v\lcx)} \bigg(
 \left[\varphi_{A}^{(2)}-\varphi_{A}^{(4)}\right](u) 
 -\int_u^1 \frac{\d w}{w} \left[\varphi_{A}^{(2)} -\varphi_{A}^{(4)}\right](w)
 \bigg) \bigg\}\; e^{-\ii u P\lcx} ,
\\
\label{Ibi_a}
\langle 0|\qb(\lcx)& \gamma_5 \ii \sigma_{\alpha\beta} h_v(0)|B(v) \rangle
 ~=~2\, \ii f_B M\frac{v_{[\alpha} \lcx_{\beta]} }{(v\lcx)}\int_0^1\!\d u  
 \left( \varphi_{T}^{(3)}(u) -\int_u^1 \frac{\d w}{w}\;\delta\left(\frac{u}{w} \right) 
 \varphi_{T}^{(3)}(w) \right)  e^{-\ii u P\lcx}.
\end{align}

The corresponding nonlocal twist-decomposed three-particle distribution amplitudes 
are given by
\begin{align}
\label{Itri_v}
\langle 0|\qb(\lcx)F_{\mu\nu}(\vartheta \lcx)\lcx^\nu &\gamma_5 h_v(0)|B(v) \rangle
%\nonumber \\ &
~=~ \ii f_B M \,\lcx_\mu
%\nonumber \\&\qquad\qquad\qquad\qquad\times
\int_0^1\!\D u_i %\left(
\Upsilon_{P}^{(5)}(u_i)
% -\int_{u_1}^1\!\!\frac{\d w}{w}\;\delta\left(\frac{u_1}{w} \right)
% \Upsilon_{P}^{(5)}(w,u_2)\right) 
e^{-\ii(u_1 + \vartheta u_2) P\lcx},
\\
\label{Itri_t2}
 \langle 0|\qb(\lcx)F_{\mu\nu}(\vartheta \lcx) \lcx^\nu &\gamma_5 \gamma_\alpha
h_v(0)|B(v) \rangle  %\nonumber \\&
~=~ \ii f_B M \int_0^1\!\D u_i \,\bigg\{\! \left(v_\mu \lcx_\alpha - g_{\mu\alpha} (v\lcx)\right) 
\Upsilon_{A1}^{(4)}(u_i) 
- \lcx_\mu v_\alpha  \Upsilon_{A2}^{(4)}(u_i)
\nonumber \\
&\qquad\quad - \frac{\lcx_\mu \lcx_\alpha}{2(v\lcx)}
\bigg( 
\left[\Upsilon_{A1}^{(4)} -\Upsilon_{A1}^{(6)} - \Upsilon_{A2}^{4)} + \Upsilon_{A2}^{(6)}
\right](u_i) \nonumber\\ &\qquad\qquad\qquad
-\int_{u_1}^1\!\!\frac{\d {w}}{w} 
\left[\Upsilon_{A1}^{(4)} -\Upsilon_{A1}^{(6)} - \Upsilon_{A2}^{4)} + \Upsilon_{A2}^{(6)}
\right](w,u_2) \bigg)%\nonumber \\
\bigg\}\;e^{-\ii (u_1+\vartheta u_2) P\lcx},%
\\
\label{Itri_t2d}
\langle 0|\qb(\lcx)F_{\mu\nu}(\vartheta \lcx)\lcx^\nu &\gamma_\alpha
h_v(0)|B(v) \rangle
~=~ -\, f_B M \;
\epsilon_{\mu\alpha\kappa\lambda}\,v^\kappa \lcx^\lambda 
\nonumber  \\&
\qquad \qquad\times
\int_0^1\!\D u_i  \left(\Upsilon_{V}^{(4)}(u_i)\; 
 -\int_{u_1}^1\!\!\frac{\d w}{w}\;\delta\left(\frac{u_1}{w} \right)\Upsilon_{V}^{(4)}(w,u_2)\right) 
e^{-\ii (u_1+\vartheta u_2) P\lcx}.
\end{align}
The nonlocal DAs corresponding to the tensor case (\ref{Dtri_g_a}) are missing here.

Since, up to now, we did not apply the on-shell constraint the results (\ref{Ibi_s}) 
-- (\ref{Itri_t2d}) can be compared to the well-known DAs for the $\pi$-meson 
\cite{Braun1990}. Of course, up to mass factors which we did not consider,
the lowest twist DAs coincide. However, due to the factors $n/(n+1)$ for the moments of
higher twist, Wandzura-Wilczek-like combinations and, due to missing zeroth moments 
analogous combinations occur. Furthermore, as has been mentioned in the Introduction, 
the dynamical higher twist contributions are related to geometric twist of the same 
as well as lower order.

% and the spin symmetry as the local distribution amplitudes.
 
%\hspace*{2.0cm}
%%\newpage
\section{Relations between Distribution Amplitudes of Geometric and Dynamical Twist}
\setcounter{equation}{0} 
\label{relations}

This section is devoted to exhibit the relations between the DAs of definite 
\emph{dynamical} twist on the one hand and of definite \emph{geometric} twist 
on the other hand. Thereby we get also relations among the DAs of definite 
geometric twist due to the heavy quark limit. (Of course, these relations could
have been obtained by applying the on-shell constraint on the expressions 
(\ref{Dbi_s}), (\ref{Dbi_v}) and (\ref{Dbi_a}) in the bilocal case, and 
(\ref{Dtri_v}), (\ref{Dtri_t2}) and (\ref{Dtri_t2d}) in the trilocal case as well.)
By construction, both types of DAs coincide at leading order but, in general, differ at 
higher order. 

First, let us present these relations for the two-particle DAs in terms of Mellin 
moments by comparing expressions (\ref{K1}), (\ref{K2n}) and (\ref{K3n})
with expressions (\ref{Dbi_s}), (\ref{Dbi_v}) and (\ref{Dbi_a}), respectively: 
\begin{alignat}{2}
\Phi_{+|n}~&=~-\varphi_{A|n}^{(2)},
&\qquad\quad
%\label{rel1} %\\
%
\hbox{$\frac{1}{2}$}\left(\Phi_{+|n}-\Phi_{-|n}\right)
~&=~-\frac{n}{2(n+1)}\left(\varphi_{A|n}^{(2)}
-\varphi_{A|n}^{(4)} \right),
\label{rel2}\\
\hbox{$\frac{1}{2}$}\left(\Phi_{+|n}+\Phi_{-|n}\right)
~&=~-\varphi_{P|n}^{(3)},
&\qquad\quad
%\label{rel3} %\\
%
\hbox{$\frac{1}{2}$}\left(\Phi_{+|n}-\Phi_{-|n}\right)
~&=~\varphi_{T|n}^{(3)}.
\label{rel4}
\end{alignat}
%(\ref{rel1}) and 
Equations (\ref{rel2}) as well as %(\ref{rel3}) and 
(\ref{rel4}) lead to independent
relations for $\Phi_\pm$. Due to the on-shell constraint in the heavy quark limit, 
the matrix elements of an axial vector operator is related to the matrix elements 
of a pseudoscalar and a tensor operator. Consequently, we find:
\begin{align}
\Phi_{+|n} ~&=~-\varphi_{A|n}^{(2)}
~=~\varphi_{T|n}^{(3)}- \varphi_{P|n}^{(3)},
\label{rel5}\\
\Phi_{-|n}~&=~-\frac{1}{n+1}\left(\varphi_{A|n}^{(2)} + n \,\varphi_{A|n}^{(4)}\right)
~%\relstack{(n \ge 1)}
{=}~- \left(\varphi_{T|n}^{(3)} + \varphi_{P|n}^{(3)}\right).
\label{rel6}
\end{align}
Due to these relations, it is sufficient to consider only either the axial vector 
operator or the pseudoscalar and skew-tensor operator without losing information 
about the heavy quark Mellin moments $\Phi_{\pm|n}$. Also the relations between 
the Mellin moments $\varphi_{A|n}^{(2)}$, $\varphi_{A|n}^{(4)}$, $\varphi_{P|n}^{(3)}$ 
and $\varphi_{T|n}^{(3)}$ -- as well as their dependence on $\Phi_{\pm|n}$ -- may be 
read off quite simply.
Especially, one finds 
\begin{align}
0~&=~\Phi_{+|0} - \Phi_{-|0},\label{RR7}\\
\varphi_{A|n}^{(4)}
~&=\;-\;\Phi_{-|n} +\frac{1}{n}\left(\Phi_{+|n}-\Phi_{-|n}\right),
\qquad\qquad\qquad~\; n\ge 1, 
\label{rel7}\\
\varphi_{A|0}^{(4)}
~&=\;-\;\Phi_{-|0} +\int_0^1 \d u \ln u \big(\Phi_{+}(u)-\Phi_{-}(u)\big),
\qquad n = 0. 
\label{rel70}
\end{align}
Eq.~(\ref{RR7}) states $\Phi_{+|0} = \Phi_{-|0}$, and consequently, with (\ref{rel2}) and (\ref{rel4}), we get
$\varphi_{P|0}^{(3)}=\varphi_{A|0}^{(2)}$ and $\varphi_{T|0}^{(3)}=0$.
Expression (\ref{rel70}) is obtained by taking into account Eqs.~(\ref{RR7}) and (\ref{rel7})
and using l'Hospitals rule.

By making use of Eq.~(\ref{MellinMoments}), we are able to obtain the corresponding nonlocal expressions as follows: 
\begin{align}
\Phi_{+}(u)~&=-\;\varphi_{A}^{(2)}(u)
~=~\varphi_{T}^{(3)}(u)- \varphi_{P}^{(3)}(u),
\label{rel8}\\
\Phi_{-}(u)~&=-\;\varphi_{A}^{(4)}(u)
-\int_u^1\!\frac{\d w}{w}
\left(\varphi_{A}^{(2)}(w)- \varphi_{A}^{(4)}(w)\right) 
%\nonumber \\&
=-\left(\varphi_{T}^{(3)}(u)+ \varphi_{P}^{(3)}(u)\right).
\label{rel9}%\\
\end{align}
The last relation looks like a (geometric) Wandzura-Wilczek relation but with missing 
non-integrated leading-twist term or, stated otherwise, the difference 
$\Phi_{+}- \Phi_{-}$ is a pure WW-term in the difference
$\varphi_{A}^{(2)}- \varphi_{A}^{(4)}$,
\begin{align}
\Phi_{+}(u)- \Phi_{-}(u) = - \left(\varphi_{A}^{(2)}(u)- \varphi_{A}^{(4)}(u)\right)
+ \int_u^1\!\frac{\d w}{w} \left(\varphi_{A}^{(2)}(w)- \varphi_{A}^{(4)}(w)\right).
\end{align}
Equivalently, from the vanishing of their 
zeroth moments we get some Burkhardt-Cottingham-like sum rules:
\begin{align}
&\int_0^1\!\d u\; \Phi_{-}(u) = \int_0^1\!\d u\; \Phi_{+}(u)\,,
\label{01}\\
&\int_0^1\!\d u\; \varphi_{A}^{(2)}(u) = \int_0^1\!\d u\; \varphi_{P}^{(3)}(u)\,,
\qquad
\int_0^1\!\d u\; \varphi_{T}^{(3)}(u) = 0\,.
\label{02}
\end{align}
The representation of the DAs of geometric twist by those of dynamic twist
reads (omitting trivial ones),
\begin{align}
\varphi_{P}^{(3)}(u)~&=~-\;\hbox{\Large$\frac{1}{2}$}\Big(\Phi_{+}(u)+\Phi_{-}(u)\Big),
\label{rel10}\\
\varphi_{T}^{(3)}(u)~&=~\;\hbox{\Large$\frac{1}{2}$}\Big(\Phi_{+}(u)-\Phi_{-}(u)\Big),
\label{rel11}\\
\varphi_{A}^{(4)}(u)~&=~-\Phi_{-}(u)
+\frac{1}{u}\int_u^1\!\d w \,\Big(\Phi_{+}(w)-\Phi_{-}(w)\Big).
\label{rel12}
\end{align}
Integrating both sides of (\ref{rel12}) over the range $0 \le u \le 1$ and observing 
(\ref{01}) the result (\ref{rel70}) is re-obtained.
%\smallskip

Now, let us present the relations between three-particle DAs of definite dynamical and geometric twist in terms of double Mellin moments by comparing expressions 
(\ref{K4n}), (\ref{K5n}) and (\ref{K2}) with expressions 
(\ref{Dtri_v}), (\ref{Dtri_t2}) and (\ref{Dtri_t2d}), respectively. 
In case of the third order tensor structure we have no projection
operator as required in the expression (\ref{tri_t3}) at our disposal. Therefore we
are unable to present the full set of relations between DAs of dynamical and 
of geometric twist. But, according to the definition of the DAs of geometric twist
we know at least that relation which results from the identity, $\delta^{(\mu'}_{(\mu}\delta^{[\alpha')}_{[\alpha)}\delta^{\beta']}_{\beta]}$, 
of the projection operator
$\widetilde{\cal P}^{(\tau)(\mu^\prime[\alpha^\prime)\beta^\prime]}_{(\mu[\alpha)\beta]|n}$, which has to be compared with the corresponding expression in Eq.~(\ref{K6n}).

Thereby, we finally obtain the following relations:
\begin{align} 
\Psi_{V|{n}}(\vartheta)~&=~\Upsilon_{V|{n}}^{(4)}(\vartheta)\,,
\label{rel41}\\
\Psi_{A|{n}}(\vartheta)~&=~\Upsilon_{A1|{n}}^{(4)}(\vartheta)
%\nonumber\\~&
~=~\Upsilon_{V|{n}}^{(4)}(\vartheta)\, + \,\Upsilon_{T|{n}}^{(5)}(\vartheta)\,,
\label{rel40}\\
X_{A|{n}}(\vartheta)~&=~
\Upsilon_{A2|{n}}^{(4)}(\vartheta)
%\nonumber \\&
~=\,-\Upsilon_{P|{n}}^{(5)}(\vartheta) 
+ \hbox{\Large$\frac{n}{2(n+1)}$}\,
\left[\Upsilon_{A2|{n}}^{(4)}-\Upsilon_{A1|{n}}^{(4)}
-\Upsilon_{A2|{n}}^{(6)} +\Upsilon_{A1|{n}}^{(6)}\right](\vartheta)\,,
\label{rel42}\\
Y_{A|{n}}(\vartheta)
~&=~\Upsilon_{A2|{n}}^{(4)}(\vartheta)\, +\,\Upsilon_{P|{n}}^{(5)}(\vartheta)
~=~\hbox{\Large$\frac{n}{2(n+1)}$}\,
\left[\Upsilon_{A2|{n}}^{(4)}-\Upsilon_{A1|{n}}^{(4)}
-\Upsilon_{A2|{n}}^{(6)} +\Upsilon_{A1|{n}}^{(6)}\right](\vartheta)\,.
%\nonumber \\&
\label{rel43}
\end{align}
Looking at Eq.~(\ref{K6n}) we observe that at most two relations connecting 
$\Psi_{V|{n}}(\vartheta)$ and $Y_{A|{n}}(\vartheta)$ with some  higher twist contributions 
$\Upsilon_{T|{n}}^{(\tau)}(\vartheta),\,\tau \ge 5,$ are missing.
Inverting Eqs.~(\ref{rel41}) -- (\ref{rel43}) the LCDAs of geometric twist 
are simply expressed in terms of LCDAs of dynamical twist. Thereby, only relations
(\ref{rel42}) and (\ref{rel43}) are nontrivial leading in the same manner as for 
the inversion of relation (\ref{rel6}) to vanishing zeroth moments: 
\begin{align}
\hbox{\Large$\frac{1}{2}$}\,
\Big(\Upsilon_{A1|{n}}^{(6)}(\vartheta)-\Upsilon_{A2|{n}}^{(6)}(\vartheta)\Big)
~&=~\hbox{\Large$\frac{1}{2}$}\,
\Big(\Psi_{A|{n}}(\vartheta)-X_{A|{n}}(\vartheta)\Big)
+\hbox{\Large$\frac{n+1}{n}$}\;Y_{A|{n}}(\vartheta),
\qquad n\ge1.
\label{rel47}
\end{align}

The nonlocal expressions are again obtained by re-converting the double moments 
to integral expressions. Thereby, we use (\ref{DoubleMoments}) to get the following 
nonlocal expressions 
\begin{align}
\Psi_{V}(u_i)~&=~\Upsilon_{V}^{(4)}(u_i)
\label{rel51},\\
\Psi_{A}(u_i)~&=~\Upsilon_{A1}^{(4)}(u_i)
~=~ \Upsilon_{V}^{(4)}(u_i) + \Upsilon_{T}^{(5)}(u_i)\,,
\label{rel50}\\
X_{A}(u_i)~&=~\Upsilon_{A2}^{(4)}(u_i)
%\nonumber\\&
~=~
Y_{A}(u_i)-\Upsilon_{P}^{(5)}(u_i) \,,
%- \hbox{\Large$\frac{1}{2}$}
%\left[\Upsilon_{A2|{n}}^{(4)}-\Upsilon_{A1|{n}}^{(4)}
%-\Upsilon_{A2|{n}}^{(6)} +\Upsilon_{A1|{n}}^{(6)}\right](u_i)
%\nonumber \\ &\qquad
%+\hbox{\Large$\frac{1}{2}$}
%\int_{u_1}^1\!\!\frac{\d w}{w} 
%\left[\Upsilon_{A2|{n}}^{(4)}-\Upsilon_{A1|{n}}^{(4)}
%-\Upsilon_{A2|{n}}^{(6)} +\Upsilon_{A1|{n}}^{(6)}\right](w,u_2)
\label{rel52}\\
%\intertext{}%
Y_{A}(u_i)~&=~\Upsilon_{A2}^{(4)}(u_i)+\Upsilon_{P}^{(5)}(u_i)
~=~
-\hbox{\Large$\frac{1}{2}$}
\left[\Upsilon_{A2|{n}}^{(4)}-\Upsilon_{A1|{n}}^{(4)}
-\Upsilon_{A2|{n}}^{(6)} +\Upsilon_{A1|{n}}^{(6)}\right](u_i)
\nonumber \\ &\qquad \qquad\qquad \qquad\qquad \qquad
+\hbox{\Large$\frac{1}{2}$}
\int_{u_1}^1\!\!\frac{\d w}{w} 
\left[\Upsilon_{A2|{n}}^{(4)}-\Upsilon_{A1|{n}}^{(4)}
-\Upsilon_{A2|{n}}^{(6)} +\Upsilon_{A1|{n}}^{(6)}\right](w,u_2)\,,
\label{rel53}
\end{align}
with $Y_{A}(u_i)$ being a pure WW-term.
Omitting the trivial relations, the inversions read,
\begin{align}
\hbox{\Large$\frac{1}{2}$}\,
\Big(\Upsilon_{A1}^{(6)}(u_i)-\Upsilon_{A2}^{(6)}(u_i)\Big)
~&=~\;\hbox{\Large$\frac{1}{2}$}\,
\Big(\Psi_{A}(u_i)-X_{A}(u_i)\Big)+Y_A(u_i)
+\int_{u_1}^1\!\!\frac{\d w}{u_1} Y_A(w,u_2),\\
\Upsilon_{P}^{(5)}(u_i)~&=~Y_A(u_i)-X_A(u_i) ,\\
\Upsilon_{T}^{(5)}(u_i)~&=~\Psi_A(u_i) - \Psi_V(u_i).
\end{align}

In the same manner as for the bilocal case we obtain Burkhardt-Cottingham-like sum rules,
\begin{align}
\int_0^1\!\D u_i\; Y_{A}(u_i)~=~0\,, 
\label{03}\qquad
\int_0^1\!\D u_i\;\Big(\Upsilon_{A2}^{(4)}(u_i)+\Upsilon_{P}^{(5)}(u_i)\Big) ~=~ 0\,.
\end{align}
In addition, due to vanishing  of 
$\Omega_{T|{0}}^{(4)}(\vartheta)= 2 \Upsilon_{V|{0}}^{(4)}(\vartheta)
+\Upsilon_{A1|{0}}^{(4)}(\vartheta)+\Upsilon_{A2|{0}}^{(4)}(\vartheta)$
and $\Upsilon_{V|{0}}^{(4)}(\vartheta)$, we get
\begin{align}
\int_0^1\!\D u_i\; \Upsilon_{V}^{(4)}(u_i)\,=\,0\,,
\qquad
\int_0^1\!\D u_i\; \Omega_{T}^{(4)}(u_i)\,=\,0\,,
\qquad
\int_0^1\!\D u_i \left(\Upsilon_{A1}^{(4)}(u_i)+\Upsilon_{A2}^{(4)}(u_i)\right)
\,=\,0\,;
\end{align} 
the last relation is consistent with expressions (\ref{tri_t2}) and (\ref{Dtri_t2}). 
From these relations we obtain
\begin{align}
\int_0^1\!\D u_i\; \Psi_{V}(u_i)\,=\,0\,,
\qquad
\int_0^1\!\D u_i \Big(\Psi_{A}(u_i)+X_{A}(u_i)\Big)\,=\,0\,.
\label{04}
\end{align} 
Furthermore, in complete analogy to (\ref{rel70}), from relation (\ref{rel47}) 
together with (\ref{03}) and (\ref{04}) we get, 
\begin{align}
\frac{1}{2}\int_0^1\!\D u_i \left(\Upsilon_{A1}^{(6)}(u_i)-\Upsilon_{A2}^{(6)}(u_i)\right)
&\,=\,
 \int_0^1\!\D u_i\,  \Psi_{A}(u_i) 
+ \int_0^1\!\D u_i\, \ln u_1\;Y_{A}(u_1,u_2).
\end{align} 

For the remaining LCDAs $\Omega_{P}^{(4)}(u_i), \Omega_{A}^{(3)}(u_i)$ and $\Omega_{A}^{(5)}(u_i)$, because of their similarity with the bilocal LCDAs, 
one gets relations analogous to Eqs.~(\ref{rel2}) -- (\ref{rel12}) with the $\varphi$'s 
replaced by $\Omega$'s (at higher twist).
 
%\hspace*{2.0cm}
%%\newpage
\section{Conclusion}
Two- and three-particle LCDAs of definite geometric twist have been introduced and 
discussed in the case of $B$-mesons. Comparing them with the corresponding LCDAs of 
dynamical twist \cite{Kodaira2001} we were able to derive relations between these 
different types of DAs as well as those relations which are due to the heavy quark 
limit halving the number of independent DAs. 
Some of these new relations, especially those for $\Phi_{+}- \Phi_{-}$ and 
$Y_{A}$ are of pure Wandzura-Wilczek type. In addition, since some zeroth 
Mellin moments vanish various sum rules of Burkhardt-Cottingham type appeared.
Concerning conventional LCDAs, this are the expressions (\ref{01}), (\ref{03}) and 
(\ref{04}) which, to our knowledge, up to now have not been considered in the literature.

In principle, applying the heavy quark symmetry analogous relations for heavy {\em vector} 
mesons could be obtained. However, these relations would be more complicated than 
those considered here. Concerning the two-particle LCDAs they could be obtained from 
the already known relations between geometric and dynamic twist for the $\rho$-meson \cite{Lazar2000} by applying the on-shell constraint.

The derivation of these new DAs makes use of projection operators onto local light-cone operators of definite twist, sandwiched between vacuum-to-meson matrix elements. Therefore, 
it was necessary to work with (double) Mellin moments and re-sum afterwards into nonlocal LCDAs. The twist-decomposed three-particle LCDAs for third rank tensors are
missing since the corresponding twist projectors were not available.

The projection operators onto local operators of definite twist, given here on the 
light-cone, are also known off the light-cone \cite{Joerg}. This opens the possibility
to consider, at least in principle, mass corrections analogous to an earlier study for $\rho$-meson DAs \cite{Geyer2001}. Of course, such corrections will interfere with and, 
therefore, supplement the mass expansion in HQET. Concerning $B$-mesons such corrections
could be obtained, at least partly, from an earlier consideration of $\pi$-meson DAs 
by Lazar (\cite{Lazar2002}, p. 68).

We expect that the discussion of the relations between LCDAs of dynamic and geometric 
twist improves the understanding of $B$-mesons. In addition, since the LCDAs of geometric
twist behave well-defined under renormalization, this approach may help to extend the 
study of renormalization properties of leading twist LCDA $\Phi_{+}(u)$ \cite{REN} also to the non-leading $\Phi_{-}(u)$ and three-particle ones.

In a subsequent paper, we are going to establish relations between two- and three-particle DAs forced by the equation of motion. In order to do so the knowledge of two-particle DAs off the light-cone is required. A first step in that direction has been made in Appendix A where the parametrization of vacuum-to-meson matrix elements is considered off-cone.

\acknowledgments
The authors are grateful to J.~Eilers for many useful discussions and hints, 
the introduction to the calculation via FORM and for providing them with the 
local twist projection operators in the compact notation. 

%\newpage
\renewcommand{\theequation}{\thesection.\arabic{equation}}\setcounter{equation}{0}
\begin{appendix}
%%%\input{Bconvent}
%\newpage
\section{\,Derivation of independent distribution amplitudes of 
dynamical twist in the heavy quark limit}
\label{GenAnsatz}

In this Appendix we derive, in the heavy quark limit, the linearly 
independent kinematic structures -- together with  Lorentz invariant DAs 
in $x$-space -- which are compatible with the tensor structure of the various bi- 
and tri-local light-ray operators when sandwiched between vacuum and $B$-meson state. 
Thereby, for the sake of convenience, all the Dirac structures (\ref{gamma}) 
are taken into account, considering $\gamma_5 \ii \sigma_{\alpha\beta}$ and its dual 
$\ii \sigma_{\alpha\beta}= (-\ii/2) \epsilon_{\alpha\beta\kappa\lambda}
\gamma_5 \ii \sigma^{\kappa\lambda}$ on equal footing. Due to the on-shell 
constraint (\ref{on-shell}) %, $v\sla h_v = h_v $, 
and because of the well-known 
relations on Dirac matrices we are enabled to derive all these parametrizations 
from a single general ansatz. By flavor symmetry, that parametrization 
is valid for any heavy meson. In addition, since in HQET spin symmetry relates 
pseudoscalar and vector mesons, a single ansatz is sufficient for both.

Furthermore, the general structure of matrix elements for bilocal 
operators $\langle 0 | \qb(x) \Gamma h_v(0)| B(v)\rangle$ may be obtained from
the trilocal operators
$\langle 0 | \qb(x) F_{\mu\nu}(\vartheta x) \Gamma h_v(0)| B(v)\rangle$ by
multiplying it with $v^\mu x^\nu/(vx)$, ignoring the dependence on $\vartheta x$ 
and renaming the obtained DAs.
Of course, also matrix elements built with the dual field strength can be
obtained this way. The consideration of trilocal operators is necessary for the
study of Wandzura-Wilczek relations in the framework of dynamical twist. For the 
same reason it is necessary to know as much as possible about the kinematic 
structure and the DAs off the light-cone. Therefore, we 
restrict our consideration to the light-cone only at the end.

Let us begin with the {\em trilocal operators}. In order to be able to apply the on-shell 
constraint (\ref{on-shell}) we choose to parametrize the following matrix element:
\begin{align} 
%\label{free5}
\langle 0| 
\qb(x) F_{\mu\nu}(\vartheta x)\gamma_5\gamma_\rho\ii\sigma_{\alpha\beta}h_v(0)|B(v)\rangle
\equiv
\left(\ii f_B M \right)
\langle F_{\mu\nu}\gamma_5\gamma_\rho\ii\sigma_{\alpha\beta}\rangle .
\nonumber
\end{align}
Here and in the following, matrix elements are simply represented by 
the different parts of their operators, also omitting the various variables. 
The most general ansatz for the independent kinematic structures which can be 
built for a tensor of rank five with two pairs of antisymmetric indices  reads 
\begin{align}
\langle F_{\mu\nu} \gamma_5 \gamma_\rho \ii \sigma_{\alpha\beta} \rangle
~=&~~
g_{\alpha[\mu}\, g_{\nu]\beta}\, g_{\rho\sigma}\, Z^\sigma_1
+
g_{\sigma[\mu}\, g_{\nu][\alpha}\, g_{\beta]\rho}\, Z^\sigma_2
%\nonumber\\&
+
g_{\sigma[\alpha}\, g_{\beta][\mu}\, g_{\nu]\rho}\,Z^\sigma_3
\phantom{\left(\frac{v^\kappa  x^\lambda}{v x}\right)}
\nonumber\\
&+
\frac{v_{[\mu } x_{\nu]} }{v x}\,
g_{\rho[\alpha}\,g_{\beta]\sigma}\,Z^\sigma_4
+ 
\frac{v_{[\alpha}  x_{\beta]} }{v x}\,
g_{\rho[\mu}\, g_{\nu]\sigma}\,Z^\sigma_5
+ 
\frac{v_{[\mu } x_{\nu]}\,v_{[\alpha} x_{\beta]} }{(v x)^2}\,
g_{\rho\sigma}\, Z^\sigma_{10}
\nonumber\\
&+
g_{\kappa[\mu}\,g_{\nu]\tau}\,
g_{\lambda[\alpha}\,{g_{\beta]}}^\tau\,g_{\rho\sigma}
\bigg(
 \frac{v^\kappa v^\lambda}{v^2}\,Z^\sigma_6
+\frac{v^\kappa x^\lambda}{v x}\,Z^\sigma_7
+\frac{x^\kappa v^\lambda}{v x}\,Z^\sigma_8
+\frac{v^2 x^\kappa x^\lambda}{(v x)^2}\,Z^\sigma_9
\bigg), 
\label{ansatz}
\end{align}
where
$Z^\sigma_i := v^\sigma X_i +  (x^\sigma/vx)\, Y_i, i = 1,\ldots,10$, with
$X_i = X_i(vx, v^2, x^2; \vartheta)$ and $Y_i = Y_i(vx, v^2, x^2; \vartheta)$ 
are altogether twenty linearly independent, Lorentz-invariant three-particle amplitudes of 
equal dimension and parity; furthermore, in this Appendix, the bracket notation 
(\ref{bracket}) is used {\em without} the factor $1/2$. 
%In the following, we omit $v^2= 1$ everywhere.

Due to the on-shell constraint (\ref{on-shell}) and the well-known identities of 
gamma matrices, the various matrix elements for different Dirac structures 
can be related to (\ref{ansatz}). In particular, it holds
 \begin{align}
 \langle F_{\mu\nu}  \gamma_\rho \ii \sigma_{\alpha\beta} \rangle~=~&   
({\ii}/{2})\, \epsilon_{\alpha\beta \sigma\tau} 
\langle F_{\mu\nu} \gamma_5 \gamma_\rho \ii \sigma^{\sigma\tau} \rangle
\label{trilocal3},\\
 \langle F_{\mu\nu} \gamma_5 \gamma_\alpha \rangle  ~=~&  ({\ii}/{6})\,\epsilon_{\alpha\rho\sigma\tau}
 \langle F_{\mu\nu}  \gamma^\rho \ii \sigma^{\sigma\tau} \rangle
 \label{P1},\\
 \langle F_{\mu\nu} \gamma_5 \rangle~=~&v^\alpha \langle F_{\mu\nu} \gamma_5 \gamma_\alpha \rangle,\\
\langle F_{\mu\nu} \gamma_5 \ii \sigma_{\alpha\beta} \rangle 
 ~=~&  v^\rho \langle F_{\mu\nu} \gamma_5 \gamma_\rho \ii \sigma_{\alpha\beta} 
\rangle
 +  2 \left(v_\alpha \langle F_{\mu\nu} \gamma_5 \gamma_\beta \rangle
       - v_\beta \langle F_{\mu\nu} \gamma_5 \gamma_\alpha \rangle \right),\\
\langle F_{\mu\nu} \gamma_\alpha \rangle
 ~=~&  ({\ii}/{6})\,\epsilon_{\alpha\rho\sigma\tau}
 \langle F_{\mu\nu} \gamma_5 \gamma^\rho \ii \sigma^{\sigma\tau} \rangle,\\
\langle F_{\mu\nu} \rangle
 ~=~&  v^\alpha \langle F_{\mu\nu} \gamma_\alpha \rangle,\\
\langle F_{\mu\nu} \ii \sigma_{\alpha\beta}\rangle
 ~=~&  v^\rho \langle F_{\mu\nu} \gamma_\rho \ii \sigma_{\alpha\beta} \rangle
 + 2 \left(v_\alpha \langle F_{\mu\nu} \gamma_\beta \rangle
       - v_\beta \langle F_{\mu\nu}  \gamma_\alpha \rangle \right).
\label{trilocal4}
\end{align}
Relation (\ref{trilocal3}) avoids the introduction of dual amplitudes and by 
equations (\ref{P1}) -- (\ref{trilocal4}) the parametrizations for all the 
basic Dirac structures (\ref{gamma})  can be derived from the general ansatz 
(\ref{ansatz}). 

Up to now, these parametrizations do not respect those dependencies 
 due to the identities for the gamma matrices.
In fact, it suffices to require the Chisholm identity, 
\begin{align}
\gamma^\mu\gamma^\nu\gamma^\alpha= \left(g^{\mu\nu}g^{\alpha\beta}
-g^{\mu\alpha}g^{\nu\beta}+g^{\mu\beta}g^{\nu\alpha} \right) \gamma_\beta
+ \ii \epsilon^{\mu\nu\alpha\beta} \gamma_5 \gamma_\beta\,,
\nonumber
\end{align} 
for the LHS of (\ref{ansatz}), 
\begin{align}
\langle F_{\mu\nu}\gamma_5\gamma_\rho\ii\sigma_{\alpha\beta}\rangle
%~=&~-\hbox{$\frac{1}{2}$}\langle F_{\mu\nu}\gamma_5\gamma_\rho
%\left(\gamma_\alpha\gamma_\beta -\gamma_\beta\gamma_\alpha \right)\rangle
~=~
  g_{\beta\rho}\langle F_{\mu\nu}\gamma_5\gamma_\alpha \rangle
- g_{\alpha\rho}\langle F_{\mu\nu}\gamma_5\gamma_\beta \rangle
-\ii \epsilon_{\rho\alpha\beta\sigma}\langle F_{\mu\nu}\gamma^\sigma\rangle\,. 
\label{rhsansatz}
\end{align}
This requirement results in a linear system of 
algebraic equations for the DAs $X_i,\,Y_i$ with unique solution:
\begin{align}
\nonumber
X_3&=X_1;\quad Y_3=Y_1;\quad Y_6=-X_7=X_5, \quad Y_8=-X_9=Y_5,
\\
X_6&= Y_7= X_8= Y_9= X_{10}= Y_{10}=0;
\end{align}
whereas the remaining eight DAs 
$X_1,~ Y_1,~ X_2,~ Y_2,~ X_4,$ $Y_4,~ X_5,~ Y_5$  
are independent ones.

Applying these restrictions to (\ref{ansatz}), computing (\ref{trilocal3}) and the 
six matrix elements (\ref{P1}) -- (\ref{trilocal4}) one gets:
\begin{align}
\langle F_{\mu\nu} \gamma_5 \rangle
~=~&
-v_{[\mu}g_{\nu]\sigma} Z^\sigma_2
-\frac{1}{vx}v_{[\mu}x_{\nu]}v_\sigma Z^\sigma_4
\label{FP}
\\
\langle F_{\mu\nu} \gamma_5 \gamma_\alpha \rangle
~=~&
- g_{\alpha[\mu} g_{\nu]\sigma} Z_2^\sigma
- \frac{1}{vx} v_{[\mu} x_{\nu]} g_{\alpha\sigma} Z_4^\sigma
%---------------------------------------
\\
\langle F_{\mu\nu}\gamma_5 \ii \sigma_{\alpha\beta} \rangle
~=~&
v^{\rho} \epsilon_{\rho\alpha\beta\tau}
\epsilon^\tau_{\phantom{\tau}\mu\nu\sigma} \; Z_1^\sigma
-
v_{[\alpha}g_{\beta][\mu}g_{\nu]\sigma} \; Z_2^\sigma
\nonumber\\
-&\frac{1}{vx} v_{[\mu} x_{\nu]}
 v_{[\alpha}g_{\beta]\sigma} \; Z_4^\sigma
+ \frac{1}{vx} v^\kappa x^\lambda
v^{\rho} \epsilon_{\rho\alpha\beta\tau}
{\epsilon^\tau}_{\kappa\lambda[\mu}g_{\nu]\sigma} \; Z_5^\sigma
\\
\langle F_{\mu\nu} \gamma_5 \gamma_\rho \ii \sigma_{\alpha\beta} \rangle
~=~&
\epsilon_{\rho\alpha\beta\tau}
 {\epsilon^\tau}_{\mu\nu\sigma}\, Z_1^\sigma
+ g_{\rho[\alpha}g_{\beta][\mu}g_{\nu]\sigma}\,Z_2^\sigma
\nonumber\\
+&
\frac{1}{vx}v_{[\mu}x_{\nu]}g_{\rho[\alpha}g_{\beta]\sigma}\, Z_4^\sigma
+ \frac{1}{vx} v^\kappa x^\lambda
\epsilon_{\rho\alpha\beta\tau}
{\epsilon^\tau}_{\kappa\lambda[\mu}g_{\nu]\sigma}\, Z_5^\sigma
\\ 
\intertext{}%-------------------------------
\langle F_{\mu\nu} \gamma_\rho \ii \sigma_{\alpha\beta} \rangle
~=~&
- g_{\rho[\alpha} \ii\epsilon_{\beta]\mu\nu\sigma}\, Z_1^\sigma
+ \ii\epsilon_{\rho\alpha\beta[\mu}g_{\nu]\sigma}\, Z_2^\sigma
\nonumber\\
+&
\frac{1}{vx}v_{[\mu}x_{\nu]}\ii\epsilon_{\rho\alpha\beta\sigma}\, Z_4^\sigma
- \frac{1}{vx}v^\kappa x^\lambda g_{\rho[\alpha}
\ii \epsilon_{\beta]\kappa\lambda[\mu}g_{\nu]\sigma}\, Z_5^\sigma
%\end{align}
\\
%\begin{align}
\langle F_{\mu\nu}\ii \sigma_{\alpha\beta} \rangle
~=~&
v_{[\alpha}\ii \epsilon_{\beta]\mu\nu\sigma} \; Z_1^\sigma
+
\ii v^{\rho} \epsilon_{\rho\alpha\beta[\mu}g_{\nu]\sigma} \; Z_2^\sigma
\nonumber\\
+&
\frac{1}{vx} v_{[\mu} x_{\nu]}
\ii v^{\rho} \epsilon_{\rho\alpha\beta\sigma} \; Z_4^\sigma
+ \frac{1}{vx} v^\kappa x^\lambda
\ii v_{[\alpha} \epsilon_{\beta]\kappa\lambda[\mu}g_{\nu]\sigma} \; Z_5^\sigma
\\
%------------------------------------
\langle F_{\mu\nu} \gamma_\alpha \rangle
~=~&
\ii \epsilon_{\mu\nu\alpha\sigma} Z_1^\sigma
+ \frac{1}{vx}
 \ii g_{\sigma[\mu}\epsilon_{\nu]\alpha\kappa\lambda} v^\kappa x^\lambda
 Z_5^\sigma
\\
\langle F_{\mu\nu} \rangle
~=~&
\ii\epsilon_{\mu\nu\rho\sigma} v^\rho Z^\sigma_1\,,
\label{FS}
\end{align}
i.e., one gets eight independent distribution amplitudes.
Since $v$ is dimensionless, all the matrix elements are of (mass) dimension 
$M^3$ -- two dimensions are due to $F_{\mu\nu}$ and one due to the bound state 
and, therefore, the structure functions are $\sim  M^2$. Furthermore, concerning
dynamical twist, the DAs $Y_i$ are subleading in comparison with the DAs $X_i$.

Now, let us truncate expressions (\ref{FP}) -- (\ref{FS}) with $x^\nu$. Thereby, 
$Y_1$ disappears and, effectively, only the following four combinations remain
as independent ones ($\Omega \equiv {v^2 x^2}/{(vx)^2}$):
\begin{align}
\Psi_V = X_1 - X_5 - \Omega Y_5 , \quad
\Psi_A = X_2 - \Omega Y_4 , \quad
X_A = - X_4 , \quad
Y_A = Y_2 + Y_4 \,.
\label{R0}
\end{align}
As result we obtain:
\begin{align}
\label{R1}
\langle F_{\mu\nu} x^\nu \gamma_5 \rangle
~=~&
\Big( x_\mu - \frac{x^2}{vx} v_\mu \Big)\big( Y_A - X_A  \big)
\\
\langle F_{\mu\nu} x^\nu  \gamma_5 \gamma_\alpha\rangle
~=~&
\big(v_\mu x_\alpha - (vx) g_{\mu\alpha} \big)\, \Psi_A
+
\frac{1}{vx}\big(x_\mu x_\alpha - x^2 g_{\mu\alpha} \big)\, Y_A
- 
\Big(x_\mu - \frac{x^2}{vx} v_\mu\Big) v_\alpha\, X_A
\\
\langle F_{\mu\nu} x^\nu \gamma_5 \ii\sigma_{\alpha\beta}\rangle
~=~&
\big(v_\mu x_{[\alpha} - (vx)g_{\mu[\alpha} \big)\,v_{\beta]}
\big(\Psi_V - \Psi_A \big)
%\nonumber\\-&
-\frac{1}{vx}(x_\mu x_{[\alpha} - x^2 g_{\mu[\alpha} \big)\,v_{\beta]}\, Y_A
+
g_{\mu[\alpha} x_{\beta]}\, \Psi_V
\\
\langle F_{\mu\nu} x^\nu \gamma_5 \gamma_\rho \ii \sigma_{\alpha\beta}\rangle
~=~&   
g_{\rho[\alpha} g_{\beta]_\lambda}
\big(x^\lambda v_\mu - (vx) g^\lambda_\mu \big)\, \Psi_A
+
g_{\rho[\alpha} g_{\beta]_\lambda}
\big(x^\lambda x_\mu - x^2 g^\lambda_\mu  \big) 
\frac{1}{vx}\, Y_A
\nonumber\\-&
 \epsilon_{\rho\alpha\beta\tau}
 {\epsilon^\tau}_{\mu\kappa\lambda} 
 v^\kappa x^\lambda\, \Psi_V
- 
g_{\rho[\alpha} g_{\beta]\kappa}  v^\kappa
\Big( x_\mu - \frac{x^2}{vx} v_\mu \Big) X_A  
\\
\langle F_{\mu\nu} x^\nu \gamma_\rho \ii \sigma_{\alpha\beta}\rangle
~=~&
- \ii \epsilon_{\rho\alpha\beta\lambda} 
\big(x^\lambda v_\mu - (vx) g^\lambda_\mu  \big)\, \Psi_A
-
\ii \epsilon_{\rho\alpha\beta\lambda} 
\big(x^\lambda x_\mu - x^2 g^\lambda_\mu  \big) 
\frac{1}{vx}\, Y_A
\nonumber\\
+&
\ii g_{\rho[\alpha}\epsilon_{\beta]\mu\kappa\lambda} v^\kappa x^\lambda\,
\Psi_V
+ 
\ii\epsilon_{\rho\alpha\beta\kappa} v^\kappa 
\Big( x_\mu - \frac{x^2}{vx} v_\mu \Big) X_A  
\\
\langle F_{\mu\nu} x^\nu  \ii \sigma_{\alpha\beta}\rangle
~=~&
\ii \epsilon_{\alpha\beta\kappa\lambda} v^\kappa \Big[
\big(x^\lambda v_\mu - (vx) g^\lambda_\mu  \big) 
\big(\Psi_V - \Psi_A \big)
%\nonumber\\+&
+ %\ii \epsilon_{\alpha\beta\kappa\lambda} v^\kappa
%\big( \big) 
\frac{x^\lambda x_\mu - x^2 g^\lambda_\mu }{vx}\, Y_A \Big]
-
 \ii \epsilon_{\mu\alpha\beta\lambda} x^\lambda \, \Psi_V
\\
\label{R7}
\langle F_{\mu\nu} x^\nu  \gamma_\alpha\rangle
~=~& 
\ii \epsilon_{\mu\alpha\kappa\lambda} v^\kappa x^\lambda\, \Psi_V
\\
\label{R8}
\langle F_{\mu\nu} x^\nu  \rangle
~=~& 0
\end{align}
Obviously, according to the antisymmetry of $F_{\mu\nu}$, after a further 
truncation with $x^\mu$, any expression vanishes.
In the course of the computation we have used 
\begin{align}
\ii \epsilon_{\alpha\beta\kappa\lambda} v^\kappa
\big(x^\lambda v_\mu - g^\lambda_\mu (vx) \big) 
-\ii \epsilon_{\mu\lambda\alpha\beta} x^\lambda v^2
~=~ &
-\ii \epsilon_{\alpha\beta\kappa\lambda}
\Big(g^\kappa_\mu g^\lambda_\nu g^\rho_\sigma 
    +g^\kappa_\nu g^\lambda_\sigma g^\rho_\mu
    +g^\kappa_\sigma g^\lambda_\mu g^\rho_\nu 
\Big)
x^\nu v_\rho v^\sigma\,,
\nonumber\\
\big(v_\mu x_{[\alpha} - (vx)g_{\mu[\alpha} \big)\,v_{\beta]}
+v^2 g_{\mu[\alpha} x_{\beta]}
~=~ &
g_{\alpha[\kappa} g_{\lambda]\beta}
\Big(g^\kappa_\mu g^\lambda_\nu g^\rho_\sigma 
    +g^\kappa_\nu g^\lambda_\sigma g^\rho_\mu
    +g^\kappa_\sigma g^\lambda_\mu g^\rho_\nu 
\Big)
x^\nu v_\rho v^\sigma\,,
\nonumber
\end{align}
together with the well-known relations concerning products of $\epsilon$-tensors
and its contractions, especially,
\begin{align}
\epsilon_{\rho\alpha\beta\tau} {\epsilon^\tau}_{\mu\nu\sigma   }
= g_{\rho\sigma}g_{\alpha[\mu}g_{\nu]\beta}
 + g_{\rho[\mu}g_{\nu]\,[\alpha}g_{\beta]\sigma}\,.
 \nonumber
\end{align}

Now, truncating also with $v^\mu $ the matrix elements (\ref{R1}) 
-- (\ref{R7}) simplify considerably:
\begin{align}
\label{S1}
\langle F_{\mu\nu} x^\nu v^\mu \gamma_5 \rangle
~=~&
- (vx)\,(1 - \Omega)\,\Phi_2\,,
\\
\langle F_{\mu\nu} x^\nu v^\mu \gamma_5 \gamma_\alpha\rangle
~=~&
 {x_\alpha}\, \Phi_1
- v_\alpha (vx)
\big[ \Phi_1 + (1 - \Omega)\,\Phi_2 \big]\,,
\\
\label{S3}
\langle F_{\mu\nu} x^\nu v^\mu \gamma_5 \ii\sigma_{\alpha\beta}\rangle
~=~&
(v_\alpha x_\beta - v_\beta x_\alpha)\,
\Phi_1
~ = ~
- ({\ii}/{2})\, \epsilon_{\alpha\beta\kappa\lambda}\,
\langle F_{\mu\nu} x^\nu v^\mu \ii \sigma^{\kappa\lambda}\rangle\,,
\\
\label{S4}
\langle F_{\mu\nu} x^\nu v^\mu \gamma_\alpha\rangle
~=~& 0 ~=~
\langle F_{\mu\nu} x^\nu v^\mu \rangle\,,
\\
%%%%%%%%%%%%%%%%%%%%%%%%
\intertext{with $\Phi_1 = \Psi_A + Y_A, \; \Phi_2 = X_A - Y_A$; truncating with $g^{\mu\alpha}$, we obtain}
%%%%%%%%%%%%%%%%%%%%%%%%%
\label{S5}
\langle F_{\mu\nu} x^\nu \gamma^\mu \gamma_5 \rangle
 ~=~&
(vx) \big(\Upsilon_2 + \Omega\Upsilon_3\big)\,,
\\
\langle F_{\mu\nu} x^\nu \gamma^\mu \gamma_5 \gamma_\alpha\rangle
 ~=~&
{x_\alpha}\,\Upsilon_1
%\nonumber\\
- v_\alpha (vx)\,\big[\Upsilon_1  
    - (\Upsilon_2  + \Omega \Upsilon_3)\big]\,,
\\
\label{S7}
\langle F_{\mu\nu} x^\nu \gamma^\mu \gamma_5 \ii\sigma_{\alpha\beta}\rangle
 ~=~& (v_\alpha x_\beta - v_\beta x_\alpha)
\Upsilon_1
~=~
- ({\ii}/{2}) \,\epsilon_{\alpha\beta\kappa\lambda}\,
\langle F_{\mu\nu} x^\nu \gamma^\mu \ii\sigma^{\kappa\lambda}\rangle\,,
\\
\label{S8}
\langle F_{\mu\nu} x^\nu \gamma^\mu \gamma_\alpha\rangle
~=~& 0
~=~ 
\langle F_{\mu\nu} x^\nu \gamma^\mu \rangle\,,
\end{align}
with $\Upsilon_1 = 2 \Psi_V  + \Psi_A + X_A ,\;\Upsilon_2 = 3 \Psi_A + X_A ,\; 
\Upsilon_3 = 3 Y_A - X_A $.

Despite of the different definitions of the DAs $\Phi_i, i = 1,2$ 
and $\Upsilon_i, i = 1,2,3$, we observe a striking similarity in the structure of the 
two kinds of matrix elements (\ref{S1}) -- (\ref{S3}) and (\ref{S5}) -- (\ref{S7}). 

Let us now consider the case of {\em bilocal operators}. The matrix elements 
$\langle \Gamma \rangle$ can be read off from the relations (\ref{S1}) -- (\ref{S4}), 
taking advantage of the similar tensor structures and assuming the two-particle 
DAs not to depend on $\vartheta$. After division by $vx$ one gets
\begin{align}
\label{S9}
\langle \gamma_5 \rangle
~=~&
- \big(\Phi_+ + \Phi_-\big)/2\,,
\\
\langle \gamma_5 \gamma_\alpha\rangle
~=~&
- v_\alpha \Phi_+ + \frac{x_\alpha}{2(vx)}\,\big(\Phi_+ - \Phi_-\big)\,,
\\
\label{S11}
\langle \gamma_5 \ii\sigma_{\alpha\beta}\rangle
~=~&
\frac{v_{\alpha} x_{\beta}- v_\beta x_\alpha}{2(vx)} \,\big(\Phi_+ - \Phi_-\big)
~ = ~
- ({\ii}/{2})\, \epsilon_{\alpha\beta\kappa\lambda}\,
\langle \ii \sigma^{\kappa\lambda}\rangle\,,
\\
\label{S12}
\langle \gamma_\alpha\rangle
~=~& 0 ~=~
\langle 1 \rangle\,,
\end{align}
where, for the sake of convenience, we introduced 
$\Phi_\pm (vx, v^2, x^2)= \big((1-\Omega)\Phi_2 \pm \Phi_1\big) (vx, v^2, x^2)$,
i.e.,~the additional $x^2$-dependence from $\Omega = v^2 x^2 /(vx)^2$
has been included into the definition of the new DAs.

Now, let us restrict onto the light-cone, $x^2 \rightarrow \lcx^2=0$. Then the  
linearly independent LCDAs, which do not depend on $x^2$, will
be written with a ``hat'', i.e., 
${\hat X}(v\lcx; \vartheta),\,{\hat Y}(v\lcx; \vartheta)$, and so on.
The kinematical coefficients of relations (\ref{R1}) -- (\ref{R8}) are shortened, 
and the distribution amplitudes are reduced as follows:
\begin{align}
&\hat\Psi_V = \hat X_1 - \hat X_5,\quad
\hat\Psi_A = \hat X_2,\quad
\hat X_A = -\hat X_4,\quad
\hat Y_A = \hat Y_2 + \hat Y_4\,,
\\
&\hat\Phi_1 = \hat\Psi_A + \hat Y_A, \quad 
\hat\Phi_2 = \hat X_A - \hat Y_A\,,
\\
&\hat\Upsilon_1 = 2 \hat\Psi_V  + \hat\Psi_A + \hat X_A ,\quad
\hat\Upsilon_2 = 3 \hat\Psi_A +  \hat X_A ,\quad 
\hat\Upsilon_3 = 3 \hat Y_A - \hat X_A\,,
\label{Upsilon}
\\
&\hat\Phi_+ = \hat X_2 - \hat X_4,\quad
\hat\Phi_- = \hat X_2 + \hat X_4 + 2(\hat Y_2 + \hat Y_4)\,.
\end{align}

Making use of these conventions and the restriction to the LC, the above derived
representations of the various matrix elements completely coincide with those 
obtained by Kawamura et al. \cite{Kodaira2001} using the trace formalism, cf.
Eqs.~(\ref{TRI1}) -- (\ref{TRI6}) and (\ref{ff1}) -- (\ref{ff3})
in the trilocal and bilocal case, respectively (cf.~Subsection \ref{parametrization}). 
Hence, we named the invariant DAs already in accordance with 
that Reference.

\section{On-cone projection operators onto geometric twist}
\label{projectors}

In Sect.~\ref{ParaGeom} we made use of (local) projections onto LC tensor operators
of well-defined geometric twist. The general procedure of decomposing non-local
$QCD$ tensor operators, either on-cone or off-cone, into a sum of such operators 
having definite geometric twist has been developed in a series of papers \cite{Geyer1999,Geyer2000b,Lazar2002,Eilers2003,Joerg}. There, it has been shown that 
this  twist decomposition crucially depends on the tensorial structure of the operator
under consideration and that it can be obtained by using appropriate projection 
operators. That procedure makes use of the representation of non-local tensor 
operators into a series of local ones and the decomposition of local tensor operators
w.r.t.~irreducible representations of the Lorentz group. The procedure simplifies 
if light cone operators are under consideration. Since local and nonlocal LC expansion 
are related mutually we can freely choose if we calculate the twist decomposition in 
the local or the nonlocal representation. 

The local LC tensors which are to be decomposed into tensors of definite twist
are given, according to the representations (\ref{BL}), by
${\cal K}^{[s]a}_{\{\sigma'\}}(v,\lcx)\,(P\lcx)^n$. In the following they are 
denoted as 
\begin{align}
 N_n(x),\quad
 O_{\alpha| n}(x),\quad
 M_{[\alpha\beta]| n}(x),\quad
 M_{(\alpha\beta)| n}(x).
\end{align}
in the (pseudo)scalar, (axial) vector, antisymmetric and symmetric tensor case, respectively.
Below, we present the corresponding local LC projection operators. The LC projection 
operators for tensors of third rank are available only in the totally symmetric case,
but this is of no relevance here.

The most compact representation of the LC projections makes use of the `interior' 
derivative \cite{Bargmann1977} acting on the light-cone,
\begin{align}
\lcdi_\alpha &= \left\{(1+(x\partial)) \partial_\alpha -\hbox{$\frac{1}{2}$} x_\alpha \Box\right\}\Big|_{x=\lcx}\,,
\label{lind}
\end{align} 
which, together with $\lcx_\alpha, X := 1+(x\partial)$ and 
$X_{\alpha\beta}:=\lcx_\beta \partial_\alpha - \lcx_\alpha \partial_\beta$
spans the conformal algebra ${so}(4,2)$. In that order, these operators 
are the generators of special conformal transformations, translations, dilations 
and rotations, respectively. Especially, it holds 
\begin{gather}
[\lcdi_\alpha, \lcx_\beta] = g_{\alpha\beta} X + X_{\alpha\beta},
\\
[X, \lcx_\alpha] = \lcx_\alpha,
\qquad
[X_{\alpha\beta}, \lcx_\mu] = g_{\mu\alpha}\lcx_\beta - g_{\mu\beta}\lcx_\alpha,
\\
[X, \lcdi_\alpha] = -\,\lcdi_\alpha,
\qquad
[X_{\alpha\beta}, \lcdi_\mu] = g_{\mu\alpha}\lcdi_\beta - g_{\mu\beta}\lcdi_\alpha\,.
\end{gather} 
From this it follows
\begin{gather}
\label{c}
(X-1)\,\lcdi_{[\alpha} \lcx_{\beta]} = - \,(X+1)\, \lcx_{[\alpha} \lcdi_{\beta]}\,,
\\
\label{ac}
\lcdi_{(\alpha} \lcx_{\beta)} =  \lcx_{(\alpha} \lcdi_{\beta)} +  X g_{\alpha\beta}\,.
\end{gather} 

The various LCDAs of definite geometric twist as well as their moments are labeled by 
$\tau = \tau_0 + r$ \cite{Eilers2003}. Thereby, $\tau_0$ is that part of the twist 
which corresponds (or would correspond) to the totally symmetric tensor operator, and 
$r= 0,1,\ldots,$ labels higher order contributions due to the actual symmetry type characterizing the irreducible representations of the orthochronous Lorentz group  
which appear in the decomposition of the light-cone operators. In fact, for the operators
$\bar{\psi}_1(\lcx)\Gamma\psi_2(0)$ with $\Gamma = 1,\gamma_\alpha,\ii\sigma_{\alpha\beta}$ 
we obtain $\tau_0= 3, 2, 1+1$, respectively, namely, if the minimal twist of $M_{(\alpha\beta)}$ is $\tau_0$ then the minimal twist of $M_{[\alpha\beta]}$ is $\tau_0 + 1$. 
In principle, there may occur different LCDAs of the same
twist $\tau$ accompanying equal kinematical structures. This takes place for tensors
of higher rank but will not be made explicit here, i.e., only their sum will be given.

Now, we state all generic non-vanishing local on-cone operators of definite geometric twist 
up to tensors of second rank \cite{Geyer2000,Joerg}: 
\begin{align}
N^{(\tau_0)}_n(\lcx) &= N_{n}(\lcx)
\label{scal}
\end{align}
\begin{align}
\label{vec_0}
O^{(\tau_0+0)}_{\alpha| n}(\lcx) &=\hbox{\Large$\frac{1}{(n+1)^2}$}\,\lcdi_\alpha \lcx^\mu O_{\mu|n}(\lcx),
\\
\label{vec_l}
O^{(\tau_0+1)}_{\alpha| n}(\lcx)&=
\left( \delta_\alpha^\mu -\hbox{\Large$\frac{1}{(n+1)^2}$} \left(\lcdi_\alpha \lcx^\mu 
+ \lcx_\alpha \lcdi^\mu \right) \right) O_{\mu| n}(\lcx),
\\
\label{vec_2}
O^{(\tau_0+2)}_{\alpha| n}(\lcx) &=\hbox{\Large$\frac{1}{(n+1)^2}$}\,\lcx_\alpha\lcdi^\mu O_{\mu| n}(\lcx),
\qquad\qquad\qquad\qquad\qquad\qquad\quad(n\ge1),
\end{align}
\begin{align}
\label{ten_A1}
M^{(\tau_0+1)}_{[\alpha\beta]n}(\lcx) &=-\hbox{\Large$\frac{2}{(n+1)(n+2)}$}
\lcdi_{[\alpha} \left( \delta_{\beta]}^{[\mu} -\hbox{\Large$\frac{1}{(n+2)^2}$}\,\left( \lcdi_{\beta]}\lcx^{[\mu} 
+ \lcx_{\beta]}\lcdi^{[\mu}\right) \right)\lcx^{\nu]} M_{[\mu\nu]| n}(\lcx),
\\
\label{ten_A2}
M^{(\tau_0+2)}_{[\alpha\beta]|n}(\lcx)
&=
\Big( \delta_{[\alpha}^{[\mu} \delta_{\beta]}^{\nu]}
-\hbox{\Large$\frac{4}{n^3(n+2)}$} \lcx_{[\alpha} \lcdi_{\beta]} \lcx^{[\mu}\lcdi^{\nu]}
\nonumber \\ &\quad~~
+\hbox{\Large$\frac{2}{(n+1)(n+2)}$} \lcdi_{[\alpha} \delta_{\beta]}^{[\mu} \lcx^{\nu]} 
+\hbox{\Large$\frac{2}{n(n+1)}$} \lcx_{[\alpha} \delta_{\beta]}^{[\mu} \lcdi^{\nu]}  
%+\hbox{\Large$\frac{2}{n^2(n+2)^2}$} \lcx_{[\alpha} \lcdi_{\beta]} \lcdi^{[\mu} \lcx^{\nu]} 
\Big)
 M_{[\mu\nu]|n}(\lcx),\quad(n\ge1),
\\
\label{ten_A3}
M^{(\tau_0+3)}_{[\alpha\beta]|n}(\lcx)
&=-\hbox{\Large$\frac{2}{n(n+1)}$} \lcx_{[\alpha} \left(\delta_{\beta]}^{[\mu}
- \hbox{\Large$\frac{1}{n^2}$} \left(\lcdi_{\beta]} \lcx^{[\mu} + \lcx_{\beta]} \lcdi^{[\mu} \right)
\right) \lcdi^{\nu]} \, M_{[\mu\nu]| n}(\lcx),\quad(n\ge1),
\end{align}
\begin{align}
\label{ten_S0}
M^{(\tau_0+0)}_{(\alpha\beta)|n}(\lcx) &=  \hbox{\Large$\frac{1}{(n+1)^2(n+2)^2}$} \,
\lcdi_{\alpha}\lcdi_{\beta} \lcx^{\mu}\lcx^{\nu} M_{(\mu\nu)|n}(\lcx),\\
\label{ten_S1}
M^{(\tau_0+1)}_{(\alpha\beta)|n}(\lcx) &=  \hbox{\Large$\frac{2}{n(n+1)}$}\, \lcdi_{(\alpha}  
\left(\delta_{\beta)}^{(\mu}
-\hbox{\Large$\frac{1}{(n+2)^2}$}\left( \lcdi_{\beta)}\lcx^{(\mu} + \lcx_{\beta)} \lcdi^{(\mu}\right)
\right) \lcx^{\nu)} M_{(\mu\nu)|n}(\lcx),\quad(n\ge1),\\
\label{ten_S2}
M^{(\tau_0+2)}_{(\alpha\beta)|n}(\lcx) &= \bigg\{
\delta_{(\alpha}^{(\mu} \delta_{\beta)}^{\nu)}
+\hbox{\Large$\frac{1}{n(n+1)^2(n+2)}$}\Big( \lcdi_{\alpha}\lcdi_{\beta} \lcx^{\mu}\lcx^{\nu} 
+\lcx_{\alpha}\lcx_{\beta}\lcdi^{\mu}\lcdi^{\nu}\Big)
-\hbox{\Large$\frac{2}{n(n+1)}$}\, \lcdi_{(\alpha} \delta_{\beta)}^{(\mu}\lcx^{\nu)} \nonumber \\ &\qquad
-\hbox{\Large$\frac{2}{(n+1)(n+2)}$}\,\lcx_{(\alpha} \delta_{\beta)}^{(\mu}\lcdi^{\nu)}
+\hbox{\Large$\frac{4}{n^2(n+2)^2}$}\,\lcx_{(\alpha} \lcdi_{\beta)} \lcx^{(\mu} \lcdi^{\nu)}
+\hbox{\Large$\frac{2}{n(n+2)^2}$} \, \lcx_{(\alpha} \lcdi_{\beta)} \delta^{\mu\nu} 
\nonumber \\ &\qquad
+ \hbox{\Large$\frac{2}{n(n+2)^2}$}\, \delta_{\alpha\beta} \lcx^{(\mu} \lcdi^{\nu)} 
+\hbox{\Large$\frac{n+1 }{n(n+2)^2}$}\,  \delta_{\alpha\beta} \delta^{\mu\nu}
\bigg\}  M_{(\mu\nu)|n}(\lcx),\qquad\qquad(n\ge1),\\
\label{ten_S3}
M^{(\tau_0+3)}_{(\alpha\beta)|n}(\lcx) &= 
\hbox{\Large$\frac{2}{(n+1)(n+2)}$}\, \lcx_{(\alpha} 
\left(\delta_{\beta)}^{(\mu} - \hbox{\Large$\frac{1}{n^2}$}
\left(\lcdi_{\beta)} \lcx^{(\mu} +\lcx_{\beta)} \lcdi^{(\mu}\right)\right) \lcdi^{\nu)} M_{(\mu\nu)|n}(\lcx),
\quad(n\ge1),\\
\label{ten_S4}
M^{(\tau_0+4)}_{(\alpha\beta)|n}(\lcx) &= 
\hbox{\Large$\frac{1}{n^2(n+1)^2}$}\,  \lcx_{\alpha}\lcx_{\beta} \lcdi^{\mu}\lcdi^{\nu}
M_{(\mu\nu)|n}(\lcx),\qquad\qquad\qquad\qquad\qquad\qquad\qquad~(n\ge2).
\end{align}
Let us remark that the restrictions in $n$, appearing in Eqs.~(\ref{ten_A2}) and 
(\ref{ten_A3}) as well as Eqs.~(\ref{ten_S1}) -- (\ref{ten_S4}) are automatically 
fulfilled due to the definitions of these expressions, i.e., the zeroth and first
moments, respectively, vanish by construction. Furthermore, the second term in the RHS
of Eq.~(\ref{ten_A1}) and the last term of Eq.~(\ref{ten_A3}) vanish -- they are
written only because of analogous terms in case of the symmetric tensor.
Finally, we should mention that 
$M^{(\tau_0+2)}_{[\alpha\beta]|n}(\lcx)$ contains two and 
$M^{(\tau_0+2)}_{(\alpha\beta)|n}(\lcx)$ contains five 
independent components corresponding to irreducible representations of the
Lorentz group which, in principle, could be accompanied by independent DAs.
But in that paper we associate only one and the same with them. Any other 
expression corresponds only to a single irreducible representation.

Obviously, these operators of definite twist are obtained by applying the 
corresponding projection operators 
$\widetilde{\cal P}^{(\tau)}_n,\,\widetilde{\cal P}_{\alpha|n}^{(\tau)\mu},\,
\widetilde{\cal P}_{[\alpha\beta]|n}^{(\tau)[\mu\nu]},\,\widetilde{\cal P}_{(\alpha\beta)|n}^{(\tau)(\mu\nu)}$ (including the fractions in $n$)
on the undecomposed operators: 
\begin{align}
N^{(\tau)}_n(\lcx) &= \left(\widetilde{\cal P}^{(\tau)}_n N_{n}\right) (\lcx),\\
O^{(\tau)}_{\alpha|n}(\lcx) &= \left(\widetilde{\cal P}_{\alpha|n}^{(\tau)\mu} O_{\mu|n}\right) (\lcx),\\
M^{(\tau)}_{[\alpha\beta]|n}(\lcx) &=
\left(\widetilde{\cal P}_{[\alpha\beta]|n}^{(\tau)[\mu\nu]} M_{[\mu\nu]|n}\right)(\lcx),\\
M^{(\tau)}_{(\alpha\beta)|n}(\lcx) &=
\left(\widetilde{\cal P}_{(\alpha\beta)|n}^{(\tau)(\mu\nu)}M_{(\mu\nu)|n}\right)(\lcx).
\end{align}
In addition, they obey the common property of projections: 
\begin{align}
\big(\widetilde{\cal P}^{(\tau)} \times 
\widetilde{\cal P}^{(\tau')}\big)^{\Gamma' n'}_{\Gamma n}
&= \delta^{\tau \tau'}\widetilde{\cal P}^{(\tau)\Gamma' n'}_{~\Gamma n},
\\
\sum_{\tau = \tau_{\rm min}}^{\tau_{\rm max}}\widetilde{\cal P}^{(\tau)}
&= \bf{1}
\end{align}
In order to prove these properties the conformal algebra and the relations (\ref{c}) 
and (\ref{ac}) have to be used.

\end{appendix}
%\newpage
%%%%%%%%%%%%%%%%%%%%%%%%%%%%%%%%%%%%%%%%%%%%%%%%%%%%%%%%%%%%%%%%%%%%%%%%%%
% THIS FILE CONTAINS ALL REFERENCES FOR PROJECT DIPLOM !
%%%%%%%%%%%%%%%%%%%%%%%%%%%%%%%%%%%%%%%%%%%%%%%%%%%%%%%%%%%%%%%%%%%%%%%%%%
%\newpage

\end{document}